# **Known Structure, Unknown Function**:
# An Inquiry-based Undergraduate Biochemistry Lab Course

Cynthia Gray[†], Carol W. Price[†], Christopher T. Lee[‡], Alison H. Dewald[§], Matthew A. Cline, Charles E. McAnany, Linda Columbus*, Cameron Mura*

## Author affiliations & correspondence

Department of Chemistry
University of Virginia
Charlottesville, VA 22904, USA

[†]Equally contributing authors

[‡]Current address: Department of Chemistry & Biochemistry; Univ of California San Diego; La Jolla, CA 92093, USA

[§]Current address: Department of Chemistry; Salisbury University; Salisbury, MD 21801, USA

*Correspondence can be addressed to LC or CM:
LC: 1 434 243 2123 (tel); columbus@virginia.edu
CM: 1 434 924 7824 (tel); cmura@muralab.org

## Document information

Abbreviations: 3D, three-dimensional; GM, group meeting; IMAC, immobilized metal affinity chromatography; JCSG, Joint Center for Structural Genomics; LDH, lactate dehydrogenase; MSA, multiple sequence alignment; MW, molecular weight; PDB, Protein Data Bank; POI, protein of interest; SALG, student assessment of their learning gains; TOPSAN, The Open Protein Structure Annotation Network; UVa, University of Virginia

Additional notes: The main text is accompanied by 8 figures, 2 tables and 7 items of Supplemental Information.

## Abstract

Undergraduate biochemistry lab courses often do not provide students with an authentic research experience, particularly when the express purpose of the lab is purely instructional. However, an instructional lab course that is inquiry- and research-based could simultaneously impart scientific knowledge and foster a student's research expertise and confidence. We have developed a year-long undergraduate biochemistry lab curriculum wherein students determine, via experiment and computation, the function of a protein of known 3D structure. The first half of the course is inquiry-based and modular in design; students learn general biochemical techniques while gaining preparation for research experiments in the second semester. Having learned standard biochemical methods in the first semester, students independently pursue their own (original) research projects in the second semester. This new curriculum has yielded an improvement in student performance and confidence as assessed by various metrics. To disseminate teaching resources to students and instructors alike, a freely-accessible *Biochemistry Laboratory Education* (BioLEd) resource is available at http://biochemlab.org.

## Introductory Overview

The undergraduate biochemistry laboratories at the University of Virginia (UVa) have been redesigned as inquiry-/research-based lab courses taught across two semesters (Chem4411/4421, Biological Chemistry Labs I/II). This redesign was spurred by the need to have students engage in novel research in the context of an otherwise typical undergraduate lab course. The first semester of the new curriculum is dedicated to instruction in modern biochemical concepts and methods, including computational biology, while the second semester focuses on an authentic (publication-grade) research question. Students apply the methods and concepts from the first semester to design and execute a functional assay of their protein of interest (POI) in the second semester. Each student's ultimate goal is to biochemically determine the function of their POI, for which the three-dimensional (3D) structure has been determined and a putative function bioinformatically annotated (based on structure), but for which no experimental functional data exist. The year-long course concludes with groups of students preparing a manuscript akin to a scientific paper and orally presenting a scientific poster that details their findings. If appropriate, the students' protein characterization results are disseminated as annotated entries in *The Open Protein Structure Annotation Network* (TOPSAN; http://www.topsan.org; [1]); thus far, nearly ten BioLEd POIs have been developed into new TOPSAN entries and, in ideal cases, student work has culminated in publications in the primary literature (e.g., [2]).

While the students focus on a well-defined research goal, centered on functional characterization of their protein, our goals as instructors include teaching students (i) how to design and execute their own experiments, (ii) how to analyze data critically, (iii) how to work in a group towards a common goal, and

(iv) how to communicate their work both orally and in writing (Fig 1 and below). A further aim has been to create a *modular* curriculum that can be adopted by instructors at any college or university; a modular design affords instructors the option to focus on discrete portions of the curriculum, versus wholly implementing all of it. In addition to assessment of the new curriculum, our final goal has been broad dissemination of the course materials. A freely-accessible ***Bio**chemistry **L**aboratory **Ed**ucation* (BioLEd) resource has been developed for this purpose at http://biochemlab.org (Fig 2, and below).

## Motivation

Published reports and peer-reviewed studies indicate that undergraduate science education must change from traditional, memorization-based instruction to a more experience-based form of learning [3-7]. These studies find that students who engage in inquiry-based learning develop better reasoning skills and more deeply enjoy research and laboratory work, versus students taught using traditional methods [8]. While traditional lab courses often utilize the same *conceptual learning* style [9] that is typically used in lecture courses, this instructional style rarely encourages students to be independent/critical thinkers. In short, there is a demand for robust and accessible undergraduate science education curricula that provide more *experience-based* learning, in a more active environment. To stimulate student autonomy and independence, such as would be required in a 'real' research environment, lab courses must focus on teaching more *procedural knowledge* [9] — including laboratory skills, experimental design, and data analysis and interpretation. Our new BioLEd curriculum employs both conceptual and procedural learning via guided instruction in the first semester (in the form of basic concepts, tools and protocols), and self– and peer–driven learning in the second semester (in the form of open-ended experimental investigations, using the concepts, tools and methods from the first semester).

Historically, UVa's Biological Chemistry Lab courses had been taught in a conventional format: structured, single-session lab exercises focused on teaching one technique via a procedure that had been validated by countless generations of prior classes. In Spring 2009, the traditional laboratory format for the second semester (Biochemistry Lab II) was abandoned in favor of a research-based curriculum, taking as a starting point the myriad proteins of unknown function that have been structurally characterized in the past decade via structural genomics initiatives [10-12]. Successive course modifications and adjustments to the curriculum followed, leading to our current year-long course design. As the course has evolved, BioLEd's curricular design and logistics have been refined in accord with national calls for changes in undergraduate science education [3, 13, 14, 5]. Across all iterations and refinement cycles, the goals of the lab course continue to be the same: to develop students' critical thinking skills, via hands-on research, and to train them in methods used in the biochemistry workforce (Table I). Students in the BioLEd curriculum engage in research from the point of inception onward. In the opening weeks of the first term, they learn bioinformatic methods and literature tools to enable them to formulate questions and hypotheses about their assigned target POI and its potential function (including what the word 'function' can mean in various contexts). Throughout the second semester, students design and execute experimental plans; they perform enzyme kinetics studies and other experiments; they collect, process and interpret data; and they communicate their findings in oral and written form (their end-of-term manuscript is in the style of a scientific research article). The role of the instructors in this course, particularly during the research-intensive second term, is to provide guidance and serve as a resource, and not to dictate the research steps directly. Here, we describe the new BioLEd initiative, which is inquiry-based (first term) and research-based (second term). We have defined precise learning gains for our modules (Fig 1 and Supp Info 1) in order to guide curricular design and refinement, and to assess student performance.

BioLEd builds upon educational principles and best practices gleaned from other efforts over the past decade. For example, the merits of a modular approach have been recognized [15], as have the benefits of group-based learning [16] and the necessity of introducing computational approaches into undergraduate biochemistry and molecular biology curricula [17]. Also, other lab curricula that utilize both the expository- and inquiry-based approaches have been recently developed (e.g., [18]), and we are not alone in suggesting protein functional annotation as a means by which to introduce undergraduates to research [19]. Appealing features of the BioLEd curriculum include: (i) its *functional genomics* framework, which leverages established biochemical methods to pursue open research questions of each POI's function; (ii) its fundamentally *modular* and *transferable* curriculum design, enabling facile adoption by other institutions/instructors; (iii) its *active learning* approaches, which pervade every aspect of the curriculum; (iv) its inclusion of *computational biology*, both informatics-based and molecular (e.g., docking).

## Description of the Course

Throughout the year-long course, students are charged with purifying and characterizing a protein for which the crystal structure was determined by the Joint Center for Structural Genomics (JCSG; [20]) and a putative function was annotated but never experimentally investigated. In order to optimize the chances of success and orchestrate course logistics, the experienced (PhD-level) instructors select proteins of interest (POI) with presumed *enzymatic* functions and assign these to students (see the *Target Selection* section for sample criteria). Students learn a wide variety of techniques to study their assigned POI in the first semester, including bioinformatic and computational methods, extensive literature surveys, and laboratory experiments in which they over-express, purify and quantify their recombinant POI. In addition, students learn how to determine enzyme kinetics via spectrophotometric assays, using the well-characterized and commercially available enzyme lactate dehydrogenase (LDH). Apart from the LDH as-say, each experiment represents truly unique and original research because each student POI group (Fig 3) is working with a different, hitherto unexplored protein.

Students continuing in the second semester of the lab are already familiar with the techniques needed to study their POI. And, because of their literature mining and bioinformatic work, they possess much background knowledge about their unique protein. This preparation allows the second semester to be less rigidly structured than the first, which is also a necessity because each POI is unique; as is true of any scientific research, a "one size fits all" approach is not feasible across the entire class. While there is a timeline for the second semester to help guide the students (Table II), much of the scientific discovery is driven directly by the students and their investment in discovering the function of their POI.

### First semester modules and assignments

Modularity and flexibility were major aims of our curricular design, such that the research and/or instructional components of BioLEd can be implemented equally easily at predominantly undergraduate institutions or PhD-granting research (R1) universities. In other words, we sought to create a course that could be comprehensive, but also amenable to only partial implementation—e.g., at institutions which do not devote a full semester or two to protein biochemistry, or if instructors wish to implement only portions of the curriculum. By creating a modular curriculum, instructors at any institution can choose to incorporate discrete elements of BioLEd into their preexisting courses.

The first-semester curriculum consists of seven experimental and five computational/discussion-based modules (Table I). Students work with their assigned POI for all modules except those involving LDH kinetics (Modules 4, 5). Modules 1a and 2 are designed to introduce students to the literature and online/web resources, and to guide them in finding articles in both the primary and secondary (review) literature that may be relevant to investigating their POI. Using a modification of the C.R.E.A.T.E. method [21], students are guided in reading and critically analyzing research articles related to their POI.

Modules 1b (pipetting) and 3 (buffers) are fairly basic types of lab activities, and students are provided with intentionally brief experimental descriptions rather than detailed protocols or specific instructions for a given task. For instance, students might be instructed to "*prepare 100 mL of 1 M Tris pH 8*", versus "*to prepare a 10x stock of Tris buffer, begin by adding 121.14 g of Tris to a clean beaker and…*". In our experience, for many students this may be their first experience with stock solutions and careful buffer/solution calculations. Students determine the detailed protocols for making the solutions they need, and they are individually tasked with making any necessary calculations as a pre-lab assignment. This approach helps instill the self-reliance and proficiency that becomes increasingly important in later stages of the BioLEd curriculum.

Modules 4 and 5 focus on kinetics assays using the enzyme LDH. This activity prepares students for the second-semester experiments, where they conduct kinetic assays on their own POIs. Module 4 requires students to determine (i) the optimal concentration of LDH enzyme for detecting signal in their spectrophotometric assays, as well as (ii) a suitable range of substrate concentrations for determining LDH kinetics parameters. Students learn how to select proper substrate concentrations to enable determination of Michaelis-Menten kinetic parameters, including the initial reaction velocity ($v_0$), maximal velocity ($v_{max}$), and the Michaelis constant ($K_M$). Sometimes, a partner pair discovers that they cannot calculate reliable kinetic parameters because the substrate concentration range initially settled upon did not sufficiently span the hyperbolic $v_0$ *versus* [*substrate*] curve. An entire lab session is dedicated to processing and analyzing the kinetics data that have been acquired (from raw absorbance measurements, to progress curves, to Michaelis-Menten plots) and, in some cases, students can repeat the experiment if they realize they did not have an appropriate range of substrate concentrations.

In Modules 7–9, students learn to transform the DNA plasmid encoding their POI into chemically competent *Escherichia coli*, over-express the recombinant POI via induction with IPTG or arabinose (depending on the plasmid), harvest and then lyse their *E. coli* cell culture, and finally purify their POI using three types of chromatography (below). These key labs introduce students to the recombinant DNA technology that was used to clone the gene for their POI, as well as the methods used to over-express and purify proteins both for this and subsequent labs (e.g., second semester). A sample protein expression/purification workflow, as executed by one of our BioLEd student groups, is shown in Fig 4.

All POIs used in our course were cloned by JCSG into either pBAD-derived (pMH4) or pSpeedET bacterial expression vectors. These protein constructs feature an N′-terminal His6× tag, enabling immobilized metal affinity chromatography (IMAC) purification on a $Ni^{2+}$–charged resin. By having students purify their POIs via affinity, gel-filtration, and ion-exchange chromatographies, they can both learn these types of chromatography and also conclude—on their own—that, in general, IMAC affords the greatest purity and yield [22]. Students also learn electrophoretic protein separation (SDS-PAGE) during these modules, and use it extensively in both semesters to monitor their protein expression and purification. Students use these methods to purify their POI in Module 9, and then learn how to quantify samples in Module 10. Students are taught the bases of two common quantitation techniques (dye-binding and $A_{280}$), as well as the caveats of each approach (e.g., the possibility of spuriously high concentrations when using the Coomassie dye-binding method, if the POI has a disproportionately high fraction of basic residues relative to the calibration standards). Students learn the advantages/disadvantages of each approach, how to execute the technique, and how to analyze the data (standard curves), all while determining the concentration of their POI samples. Module 10 also leverages the dye-binding quantification method to introduce the concept of protein–ligand binding assays. Ligand-binding experiments that are tailored to each student's POI are not easily performed because (i) each group works with a unique POI, (ii) the potential ligands to each POI are unknown, and (iii) data for binding isotherms are not readily acquired (at least not with the detection methods and equipment found in most undergraduate biochemistry labs). Rather, the topic of ligand-binding equilibria is introduced by quantifying the binding of Coomassie to bovine serum albumin, as described by Sohl & Splittgerber [23]. Given suitable equipment and available materials, students may propose similar POI–specific experiments in the second semester.

The computational biology components (Modules 6 and 11) guide students in using both informatics–based (Module 6) and chemistry–based (Module 11) computational methodologies as a way to quantitatively explore the sequence/function and structure/function relationships for their POI. These modules rely on the deep *sequence* ↔ *structure* ↔ *function* paradigm at the heart of biochemistry (Fig 5). We introduce students to both families of approaches for inferring protein function: (i) the statistical/data-driven approach of bioinformatics (Module 6; Fig 5B) and (ii) the chemical/structure-based approach, as exemplified by molecular docking (Module 11; Fig 5C). A key lesson taught here is the *comparative approach* in biology: Students learn that they can use systematic comparisons at the levels of sequence and structure, between their POI and proteins of well-characterized function, to predict potential functions of their POIs (e.g., substrate specificities). Then, they design experiments to test those predictions in the second semester. Throughout these Modules, students are taught structural bioinformatics concepts and jargon ('homology', 'domain', 'superfamily', 'fold family', etc. [24]), as well as the principles of sequence-based bioinformatics (e.g., BLAST expectation values). Students learn, for instance, that being able to classify their POI into a particular fold family does not necessarily provide sufficiently detailed information to allow meaningful (*specific* and *testable*) hypotheses for a POI's substrate specificity.

In Module 6, students employ bioinformatic servers, databases, and literature-search methods to help identify potential enzymatic activities, substrate specificities, and any function-related motifs in their POI. This is done at the levels of sequence and structure (Fig 5A). This module demands a highly immersive learning approach and, because this material is new to many students, a more planned approach may be necessary at this stage (e.g., we have had students pattern their workflows after Mazumder & Vasudevan's approach [25] to structure-guided comparative analysis of protein function). We first introduce basic concepts, including PDB file manipulation [26], sequence alignments, and phylogenetic trees. We then introduce students to powerful bioinformatic tools for (i) structure comparison, both pairwise (e.g., in the PyMOL molecular visualization environment [27]) and against structural databases (e.g., VAST [28], DALI [29]); (ii) integrated structure analysis services (e.g., PDBsum [30]); (iii) comprehensive sequence/function databases such as UniProt [31]; and (iv) databases and knowledge-bases with a specific focus on enzyme function (e.g., BRENDA [32]) or pathways (e.g., KEGG [33]). This Module should emphasize to students that knowledge gleaned from database searches and analyses can be integrated with careful study of any primary literature that might be available for functional characterization of close homologs of their POI.

Module 11 introduces students to what can be learned by detailed analysis of the 3D structure of their POI. Molecular visualization [34], modeling approaches (e.g., homology modeling), and protein/ligand docking [35] form the core of this module (Fig 5C). Students examine the features of their structure using PyMOL, which they are introduced to early in the semester and which we then revisit in class (e.g., Supp Info 2). Students optionally build homology models using Swiss-Model [36], and conduct docking experiments with PatchDock [37]. We recently developed a standalone (non–web-based) educational workflow for docking that uses the high-performance AutoDock-Vina software [38]. In this workflow (Supp Info 3), students learn docking as a powerful *in silico* tool for exploring the ligand-binding properties (and hence function) of their POI, and the students also learn basic usage of the Linux operating system (this is an exciting first for many students). We have found that students need close guidance in order to learn to distinguish less relevant small molecules in a PDB file (e.g., glycerol from crystallization conditions) from more promising cofactors, metals or other ligands that might be bound, and to learn how to navigate and interpret the vast information content of a PDB file. Similarly, we find that most undergraduates must be carefully introduced to the notion of protein packing in a crystal lattice, and how such packing may relate to the biologically functional oligomeric state; this is an especially important point as regards students' structural analyses of POIs that are suspected to act as multimers.

Another lesson regarding the computational biology modules is that BioLEd is generally the first lab course encountered by biochemistry students (at least at UVa) that does not expect specific, preordained, 'right-or-wrong' answers. Indeed, we have found that a difficulty in facilitating the bioinformatic labs is that many students expect questions to have a single 'right' answer; thus, a common pitfall is that many students are tempted to mechanically 'plug and chug' data into bioinformatic servers, rather than explore, critically analyze, and ruminate about the results for their POI. Instructors and TAs can preempt this difficulty by repeatedly emphasizing that *active investigation* and *digging* (*data mining*) will yield interesting discoveries and putative leads about possible POI functions. During all computational biology sessions, the instructional staff should engage the students about their findings in 'real time'. For instance, as students are conducting sequence similarity searches for homologs, TAs can question them about the total number of 'hits' detected beyond the statistical threshold, how the number of *new* hits changes after 1, 5, 10, … iterations of PSI-BLAST [39, 40], and so on.

### Second Semester: Summary

Unlike the first-semester biochemistry lab, a pre-set syllabus of laboratory modules and associated protocols/guides is not provided to students in the second semester. Instead, the students are charged with planning their work: They formulate a strategy and timeline in consultation with the instructors. We provide a general outline of experimental progress, as an idealized plan for students to follow (Table II), but they are free to propose deviations from it. As with 'real' biochemical research, students often find that they must adapt their second-semester plans based on the outcomes of their individual experiments and the general behavior (solubility, etc.) of their POI.

Students over-express and purify their POI using the knowledge they gained in the first semester—namely, the chromatographic purification method that gave the best results with their POI. (In general, most students proceed via IMAC with their $(His)_6$–tagged POIs.) Next, the purified POI has to be exchanged into a buffer in which the protein is soluble at their working concentrations, and which is compatible with the planned enzymatic assays. The students must discover what types of buffer conditions others have used to study homologous proteins, and what conditions work with those homologs that have been confirmed as having the same enzymatic activity that the POI is thought to have. Designing this experiment requires students to use the literature skills that they developed in the first semester.

Determining a suitable buffer, both for enzyme storage and enzyme assays, can be challenging and time-consuming. General guidelines, including a discussion of the importance of salts, ionic strength, pH and protein concentration, help the students get started in selecting buffers, and also provides a starting point for troubleshooting solubility issues; nevertheless, suitable buffers typically must be determined by empirical trial-and-error. Because each POI has already successfully traversed the typical structural genomics *cloning → over-expression → purification → crystallization* pipeline, in principle students should be able to obtain high yields of pure, soluble protein for each POI target. Regardless, roughly one-quarter of our POIs over the past few years have proven exceptionally challenging, and simply obtaining conditions which allow the protein to remain soluble might be judged as being sufficient (in terms of student grades).

After obtaining pure protein in a suitable buffer, students must optimize the POI concentration that will be used in enzymatic assays throughout the semester. We define an 'optimal' amount of POI as enough protein to obtain a reliable kinetics signal, but as little protein as possible so that many assays can be performed with one preparation; in addition to maximizing throughput, this strategy reduces the variation between sample preparations. Optimizing the POI concentration requires certain concepts to be understood. The spectrophotometer 'blank' and the 'negative control' (for background rate subtraction) are especially confusing to students, partly because there is not a standard/well-defined terminology in the literature. The 'blank' for the spectrophotometer can be confused with an 'enzyme blank,' which ac-tually is more accurately considered a 'negative control,' 'blank rate,' or 'background rate.' For example, for the LDH assays we teach students that the spectrophotometer *blank* consists of all assay solution components except for the enzyme and any light-absorbing cofactors being monitored (NADH); the *blank* is used to set the absorbance of the spectrophotometer to zero. The *background rate* is the change in absorbance signal of the full assay solution—minus enzyme—over the same length of time that enzymatic activity is monitored. Establishing the background rate is important because the next step is to discern a *significant* signal versus background noise. Concepts such as the instrument's detection limit and the background signal must be thoroughly discussed in order to ensure that students can discern when their data reveal authentic activity, as opposed to data that differ only insignificantly from the background rate.

Some target POIs are almost certainly misannotated [41] or annotated at only low functional resolution in public databases. This means that the substrate(s) the students chose to test might be inappropriate for the POI, yielding negative results. Distinguishing true negative results from student error requires a positive control. However, a true positive control is impossible because the POI functions are unknown. When coupled reactions [42] are used, we provide students with the substrate of the coupling enzyme, allowing them to observe and measure the activity of the coupling enzyme alone. Finding activity for the coupling enzyme(s) alone reassures students that their reaction conditions are favorable for the planned assay. In the case of a direct assay, a commercial enzyme (if available) is used as a positive control, again providing assurance that assay conditions are compatible with enzymatic activity. In addition to planning suitable controls, students should plan to test alternative substrates in the event that their putative function is not supported; selection of viable alternative substrates can be guided by bioinformatics, docking results, and the literature.

Upon initial detection of activity and optimization of POI concentration, students determine the kinetic parameters with one substrate. Doing the experiment in triplicate to obtain standard deviations is important—students are typically intrigued by the variation they find, and they become more critical of articles in the primary literature that do not report standard deviations or other statistical estimates of error. Once students have acquired and processed the kinetic data, they are encouraged to systematically vary the assay to begin investigating the catalytic mechanism, protein stability, and/or substrate specificity. Students often choose to vary the pH, temperature, available metal cofactors, or test the effects of inhibitors that they chose based on bioinformatic analyses.

### Second Semester Assignments

This research-based curriculum involves assignments that are atypical for a standard lab course. For the first assignment of the second semester, student teams prepare lists of required chemicals and an outline of their planned experimental (kinetic) assays. Next, independently written assignments require each student to detail the materials and methods used in their work (week 6), and to write a report with introduction, figures, and future goals (week 9). Both of these assignments prepare students to write a POI manuscript that is due at the end of the semester (week 13). Students benefit by having the final, large-scale assignment consist of these sub-tasks distributed throughout the semester, and they are also able to incorporate the feedback they receive on the smaller assignments into the final manuscript. Having these assignments earlier in the semester also ensures that students are sufficiently immersed in their POI. In addition to the final manuscript, each POI group (Fig 3) creates a poster for an end-of-term poster session, simulating the experience at a scientific conference; while the poster is prepared as a group effort, individual students take turns presenting the poster to the instructors and teaching assistants. In the past few years, dozens of BioLEd students have presented their results as posters at a local meeting of the American Chemical Society; this late-April event is opportunely timed just before the end of each Spring term, and similar regional events likely can be found near other institutions considering a BioLEd-based curriculum.

Two group meetings (GM) per POI are held in the second semester, as detailed in the *Teaching Communication & Critical Thinking* section (below). These meetings mimic group meetings held in research labs, our aim being to encourage interactions among students and between students/instructors, and to train students in effective scientific communication. To help students prepare for their final poster and manuscript, instructors should provide discussion and feedback on student figure preparation, how data are presented (types of plots, etc.), and the overall quality of the GM presentation. Instructors also analyze and discuss the scientific content of these presentations, so as to fully grasp the data that students are generating as well as the overall progress of each POI project. The GMs are spaced roughly ⅓ and ⅔ of the way through the semester, giving milestones to help students remain focused on the ultimate goal of characterizing the enzyme kinetics of their POI.

### Second Semester Grading

Group-based projects, which are at the level of an entire POI group rather than individual students or partner-pairs (Fig 3), comprise a relatively large share (≈30%) of a student's final grade in the second semester. This group work includes making GM presentation slides, collaborating on the end-of-term poster, and writing the final POI manuscript. Non-group components of the second-semester grade include an individual student's performance on the GM presentations (≈15%); their weekly quizzes, notebook, and effort grades (≈8%, 7%, 10% respectively); and other individual lab reports during the term (≈30%).

Because the BioLEd curriculum is one of real research, the grades for the final project and presentations are based not on 'positive' results, but rather on criteria such as students' use of the scientific method (e.g., systematic controls), scientific inquisitiveness, problem-solving efforts, resourcefulness, and overall effort. Some of our past POI targets have been difficult *in vitro*, generally due to protein solubility issues or because no enzymatic activity was detected. In such cases, little to no kinetics data were obtained by students. A lack of 'results' from such POI groups does not insure a low grade; instead, it is made clear to students that they would be expected to (i) use the primary literature to investigate potential reasons for difficulties, (ii) develop troubleshooting scenarios, and (iii) rely more heavily on computational biology to investigate the putative function of their POI. In this way, students learn that they are engaged in very real scientific research; we have found that many students embrace these challenges.

## Infrastructure

As implemented at UVa, the BioLEd-based course meets for a one-hour class and four-hour laboratory session per week. On occasion, less formal review sessions or office-hours are also offered (e.g., a session dedicated to the Michaelis-Menten and related kinetics equations). Two research-active faculty, one full-time instructional laboratory support specialist, and six graduate student TAs serve the third- and fourth-year undergraduates enrolled in the year-long course. Notably, the BioLED approach scales well: Of our 100+ chemistry graduates per year, typically ≈80-90 have enrolled in our biochemistry lab (predominantly chemistry majors with a biochemistry focus). There are six lab sections per week, each led by one graduate TA. Each TA/lab section is assigned two POIs; each POI typically has 5-9 students, working in pairs or triples (Fig 3). In total, 12 POIs are studied each year, distributed across the six laboratory sections.

Lab sections meet in one of two laboratory spaces. Each laboratory is equipped with, for every 2-3 students, a UV/Vis spectrophotometer, a gel electrophoresis setup, and a stir plate. Each laboratory also has shaker-incubators for cell growth and protein expression, a standard centrifuge, pH meter, scales, assorted chromatography columns, and other typical biochemistry laboratory equipment.

### Teaching assistants

The course described above requires six teaching assistants; thus far, we have accommodated student:TA ratios as high as 18:1. TA preparation is vital to the success of this laboratory. Most graduate students in the department teach in their first two years. Because most of the TAs are new to research themselves, several hours are scheduled to train them before the Fall term begins. TAs are introduced to the pedagogical principles, best practices, and instructional strategies underlying the BioLEd curriculum. During the semester, TAs are expected to perform all of the computational labs and create the keys used in grading those assignments. Each lab protocol is discussed in detail (at a TA meeting near the start of each week) in order to identify thin areas in a TA's knowledge-base. Also, novice TAs who teach a Wednesday or Thursday section are encouraged to observe at least part of a lab earlier in the week. Because we find that TAs often hesitate to reveal when something is new or unfamiliar to them, the TA training module is evolving to include an actual dry-run of each laboratory technique (rather than simply a discussion).

## Content Delivery, Active Learning

The limited time for instructor–student interaction is a difficulty in implementing the BioLEd curriculum in a typical (3-credit) undergraduate biochemistry lab. In one hour of lecture per week, the instructor may seek to cover the theory of the method(s) being used in lab that week, practical aspects of implementing a method, various aspects of statistical data analysis/interpretation, and so on. All the necessary content cannot be covered in a one-hour lecture. In addition, data analysis/interpretation is more effectively learned actively, rather than by lecture. Thus, the typical lecture has been replaced with an inverted lecture style [43-45]. Lecture content was recorded as brief (<15 minutes) slideshow videos and supplemented with reading assignments. Practical execution of lab methods was also provided as videos, either found online or created in-house. The weekly lecture hour was thereby freed so the instructor could actively work through sample calculations, describe anticipated data/graphs, interpret data, demonstrate the usage of software and databases, and answer any troubleshooting questions.

Interactive teaching [46-48, 43, 49-51] is an effective tool for delivering most of the lab course content. For instance, for the lecture on ion-exchange chromatography, students are asked to draw a putative chromatogram on the board. One student volunteer might draw the axes ($A_{280}$ and ionic strength as *y*-axes), while others may make changes based on feedback from the class and instructor. Further student volunteers will then draw a typical $A_{280}$ trace and ionic-strength trace. A final student might then be asked to sketch the expected SDS-PAGE gel of specific fractions from the chromatogram; this is an especially valuable exercise for gel-filtration chromatography, where any oligomeric POI that elutes should migrate at the mass of a monomer on a denaturing gel. The class is encouraged to add or otherwise edit what is drawn on the chalkboard, and especially to ask questions. This interactive format engages students and encourages active participation. Those students not actively participating at any given moment are nevertheless thinking about what they would draw, and are able to work through their ideas via discussion. During some lecture times, the class works through problems in pairs or small groups; representatives from each group volunteer to share their answers. This format is particularly helpful for the Buffers & Solutions module (Module 3, Supp Info 1), as the concepts of stock solutions and dilutions are cemented via calculations and the practice of making the solutions.

Another active learning strategy—concept mapping—is introduced in week two (Module 2). First, the instructor shares a concept map about a topic that should be familiar to students from past coursework (e.g., hemoglobin). The instructor explains how the hemoglobin map was created, and that each concept map is unique. Another familiar topic is then chosen, and individual students begin creating a concept map on the chalkboard, connecting ideas, facts and concepts related to the new topic. The rela-tionship between the concept map and literature search keywords is easily introduced by having students combine words from the map, use these as literature search queries, and then compare the results.

Active learning can also be used to teach data analysis. A data figure can be projected, and the students can be asked questions that are either factual (e.g., *what method was used to generate the data? what controls are present/missing?*) or interpretative (e.g., *what hypothesis might these data address? what conclusions can be drawn?*). For each type of question, multiple answers are heard, compared, and discussed amongst the students and instructors (including TAs). This instructional mode is especially important in the second semester, when some student groups start generating potentially large amounts of kinetics data for their POIs. In the second semester, those lecture hours that are not scheduled for activities such as group meetings can be used to reinforce important concepts (e.g., analyzing progress curves to extract Michaelis-Menten kinetic parameters), address any recurring troubleshooting issues, etc.

All of these active and interactive learning methods have been highly effective in the BioLEd curriculum, based on our initial assessment results (described below). In general, the instructional tools and best-practices to be deployed in a specific course will vary with the exact concepts, sets of students, and instructors involved; this aspect of curriculum design should be researched by an instructor to identify what are likely to be the most suitable styles for a given course [44]. Numerous active learning options exist for teaching different types of concepts (e.g., [44] and [52]).

## Target Selection & Preparation

The proteins selected for students as target POIs generally meet certain criteria: (i) a 3D structure of the protein is available; (ii) the protein function is unknown/unreported; (iii) the putative function is likely enzymatic (as inferred from bioinformatics); (iv) the enzymatic reaction can be monitored via spectrophotometric assays (either directly or via coupling reactions); and (v) all substrates, cofactors, and coupling enzymes are commercially available and are affordable. Before the term begins, a PhD-level instructor evaluates each POI candidate against these criteria. As targets are selected, corresponding clones are requested from collaborators at the JCSG or are purchased from Arizona State University's *Plasmid Repository* (http://dnasu.asu.edu/DNASU). To verify that correct target plasmids have been obtained, and to prepare materials for the students for Module 7, plasmid DNAs are mini-prepped/sequenced by the instructional staff before the term begins.

Based on our experiences with over 40 POI targets, we recommend avoiding dehydrogenases with vague annotations (e.g., an 'alcohol dehydrogenase') unless the operon structure or other bioinformatic data strongly suggest a specific substrate. As an example, we have had a student group who surveyed over 20 substrates for one POI with no positive results, implying that 'dehydrogenase' is an insufficiently precise descriptor for this class of enzymes. Though negative outcomes do not affect the students' course grades (see above), confidence and morale can become diminished in these POI groups.

## Facilitating Group Work

Along with the call for science to be taught in a more experiential manner, there has been a call for teaching in a more collaborative and cooperative way [53]: "*The collaborative nature of scientific and technological work should be strongly reinforced by frequent group activity in the classroom. Scientists and engineers work mostly in groups… Similarly, students should gain experiences sharing responsibility for learning with each other.*" In addition to learning the skills of working within a group, students often learn and retain more when they work in small groups on projects (e.g. cooperative learning [54, 55]) versus other instructional formats [54, 56, 57, 55]. BioLEd students experience cooperative learning, the characteristics of which include (i) students working in small groups, (ii) students experiencing shared learning goals (and tasks that may differ from those of other groups), and (iii) grades that are based on both individual work and group work.

Group work can be difficult to implement, largely because of the personality and aptitude differences inherent to any collection of human beings. More than three years of experience in implementing the BioLEd curriculum reveals that many challenges directly stem from intra-group dynamics. Common issues include (i) a student feels that the workload/contributions in their group are unequal; (ii) a lack of communication, electronically and in person; and (iii) irresponsibility on the part of one group member hampers the entire group (e.g., someone forgets to come into lab to start an overnight culture, thus delaying their entire group by at least a day). These types of issues are common to cooperative learning, and can be addressed by incorporating the following practices:

i) *positive interdependence*: students learn that their success is tightly coupled to the contributions and success of others in the group

ii) *face-to-face positive interaction*: students must be encouraged to directly interact, both during discussions (such as the group meetings) and by sharing information

iii) *individual and group accountability*: students are held accountable both for their individual work and for contributing sufficiently to the group project; thus, both individual and group grades factor into the overall grade

iv) *group processing*: students are given opportunities to 'grade' their group's functionality, and to discuss what have been positive and negative aspects of their experience working in their POI group

In the first semester, students have a strong incentive (individual grades) to stay on-task and be prepared each week (quizzes, pre-labs). This intentional course structure helps reinforce student independence. In contrast, the second semester leaves preparedness and time management to the students. Also, grading methods are required that specifically address the issues associated with group work. A bal-ance of group and individual grades was found to be crucial in order for students to appreciate that their grade does incorporate their personal effort and intellect, regardless of the effort and performance level of their peers. For example, the overall grade for the second-semester GM presentations includes group and individual subtotals. Effort reports are prepared by each student and turned-in with select assignments (Supp Info 4 is a sample). These reports are vital for an instructor's evaluation of the group, and also for students to pause and consider the contributions of each group member. Though students tend to be generous with one another in scoring overall effort, students who do not contribute are easily identified by this mechanism. Questions pertaining to what each individual student brought to the group, and what the student learned from the group, help students appreciate the benefits of working together cohesively.

## Teaching Communication & Critical Thinking

### Group meetings

Group meetings occur twice in the second semester, at weeks 7 and 10. Each GM is attended by the instructors and TAs for that POI. The meeting format mimics that in most research labs. Students present a collaboratively-prepared slide presentation in a small group setting; presentations are followed by discussions. In advance, students are given an outline of what sort of information should be included in their presentation slides. To ensure that each person is familiar with all of their group's work, students are told to be prepared to present any segment of the presentation; slides are assigned to individual students at the start of the GM. These meetings are kept intentionally informal and interactive, and it is useful to bear in mind that many undergraduates will be nervous about speaking in front of their professors.

The GMs are valuable on many levels. First, science majors are rarely expected to present their work in class, and therefore they do not gain experience in articulating and defending their ideas 'on their feet'. The BioLEd curriculum affords opportunities for students to gain confidence in communicating their work via scientific/technical speaking, in a low-key and welcoming environment. Second, the GMs help instructors track the students' progress with each POI, individually and as a group. In classes with large numbers of students and sections, instructors likely will be unable to stay abreast of each POI project without such meetings. The GMs also allow for interactive brainstorming and troubleshooting. Because much of this course entails group work, a student's individual turn in presenting part of the GM is a key opportunity to demonstrate their mastery and ownership of the work (i.e., apart from the group work to prepare the slides); also, the instructors can gain a sense of how the group is functioning. An important result of the GMs is enhanced faculty-student interactions in an intimate setting. Studies indicate that such environments are especially important for novice students, whose needs differ from students with research ex-perience [58-60]. Personal interactions with research mentors can address some of the differences, by providing guidelines and orientation, as well as socialization in the traits of scientific researchers.

## Final Manuscript

The BioLEd students' research culminates in a manuscript prepared in the style of a scientific publication. Students work towards this final report throughout the semester in discrete stages, corresponding to the sections of a typical scientific article (*Introduction*, *Methods*, etc.). For the final report assignment, students merge their adapted *Materials & Methods* (from week 6) and *Introduction*, together with figures and future work (from week 9) and newly written *Abstract*, *Discussion*, and *Results* sections. Much of this final paper involves bioinformatics, which students were introduced to in the first semester and urged to revisit since then. For instance, the *Introduction* contains the students' hypothesis about the function of their POI (e.g., substrate specificity of a putative aminotransferases), which forms the starting point for the second semester's group work. Students learn that their hypotheses have to be justified, largely via bioinformatic results with their POI and by analysis of the salient literature for any homologs.

The manuscript is a group project. There are many reasons for this. First, there are several facets to the manuscript, and each student brings different strengths to bear on the research and writing; this reflects how research groups actually work in academia, national labs, industry, etc. Second, students work throughout the semester to study a single POI in groups of up to 6-8 students (Fig 3). Early on, we prod students to consider working on different aspects of the POI (i.e., as a synergistic group rather than just a collection of individuals); this way, they accomplish more than they thought possible. However, only one manuscript per POI is accepted. Thus, students are made to work together to craft their findings into a cohesive description. Achieving this goal teaches students efficient scientific communication. Finally, because peer review and critical data analysis are important skills for scientists, groups are encouraged to hash through a series of drafts and edits. We also require a breakdown of each individual's contribution to the manuscript; this accountability helps promote a fair division of labor towards the manuscript. The final manuscript may develop into a line of further work: If the final results for a given POI are definitive, demonstrating either (i) the annotated enzyme activity, (ii) absence of the predicted activity/substrate specificity or (iii) some other activity/specificity, then the instructor can work with any interested students from the POI group to draft a new annotation entry for submission to TOPSAN (see above). And, if the POI results are publishable, then instructors can recruit students from the POI group towards such efforts. At least some further experiments (beyond those in the final project report) are generally required before being able to publish, and such work can be pursued the next summer or academic year (e.g., for research credit); indeed, one recent student developed their BioLEd project into an MSc thesis in our own (research) laboratories.

### Poster Preparation and Presentation

Most undergraduates are unfamiliar with the ways scientists present and share their work at meetings. Preparing a scientific poster requires students to mine their data and results (which are reported in detail in the final manuscript), distilling the work into only the most compelling results and effective figures. In addition, students must be prepared to lead an audience through the contents of their poster. Students who do this well typically possess a deep knowledge of both their POI as well as each group members' contributions towards characterizing the POI. The poster exercise gauges student familiarity with *what* work was done as well as their grasp of *why* particular sets of experiments were pursued.

The poster presentation is also an opportunity for students to practice scientific speaking. Unlike the GMs, each student walks the instructors and TAs through the poster and explains the entire project on their own. So, while the poster is generated as a group effort, the posters are presented individually. Some students who do not do well on written work are found to shine during oral presentation, with their depth of knowledge more readily apparent in these real-time interactions. Thus, student performance is optimally assessed by not limiting the graded work to only written assignments.

## Assessment

Our approach to course assessment was multi-pronged, the overall goal being to gauge the effectiveness of the new BioLEd curriculum. We evaluated the curriculum in three ways: (i) student gains in scientific content were assessed by us via pre–/post–course tests; (ii) student performance and content gain were self-assessed with pre–/post–course surveys; and (iii) university-wide course evaluations were used. (All assessment activities were approved by the UVa Institutional Review Board for the Social & Behavioral Sciences [#2010041200] and were in compliance with their policies.) We also surveyed past students on their opinions of the course; specifically, we questioned BioLEd alumni on whether this course gave them a deeper understanding of biochemistry and enabled them to approach scientific problems more effectively.

### Assessment of Student Performance on Assignments

For purposes of both grading and course assessment, we defined the four learning gains shown in Fig 1: (i) Aims & Concepts, (ii) Experimental Design, (iii) Data Processing, and (iv) Broader Context. These four learning gains are further defined by eight focal areas: (i) laboratory skills, (ii) broad biochemical knowledge, (iii) reading/comprehending scientific articles, (iv) written and oral communication, (v) group dynamics skills, (vi) investigative skills, (vii) critical thinking, and (viii) problem-solving skills. Each learning gain and focal area can be evaluated by specific outcomes (examples are given in Fig 1). Outlining learning gains—and using detailed grading rubrics based on these gains and focal areas—are important steps in assessing student performance in a newly developed curriculum. In the first semester, the TAs grade assignments using rubrics that we developed for two purposes: to enable assessment of students in our four learning gains, and to help focus a TA's grading efforts on those concepts specific to each assignment; a sample rubric is shown in Supp Info 5. Though necessarily detailed, BioLEd rubrics cannot be *too* specific because each report may be quite unique (reflecting the properties that are unique to each POI).

By having TAs complete rubrics when grading student assignments, the scores become more reliable and consistent for students within one section and also among different sections (different POIs, different TAs). When graded assignments are returned, students can see what they did well and what areas might require improvement. Furthermore, because the rubrics are based on our learning gains, TAs and instructors can refer to the rubrics for the main focus of a given assignment. If many students seem to struggle on particular assignments, then TAs/instructors can begin to detect patterns, such as a particular learning gain that may require more attention in the classroom or laboratory. In addition to assessing student performance via well-defined assignments, we also used concept inventory tests to assess content gain and retention at the start of the first term, end of the first term, and end of the second term. The initial results of these studies (outlined below) indicate that most students in the BioLEd curriculum demonstrate sustained learning gains in almost all topics, across the entire year.

### Student Self–assessment of Learning

An assessment mechanism using pre- and post-course surveys, created with the web-based *Student Assessment of their Learning Gains* (SALG) program [61], was used to examine student confidence levels and self-reported learning gains. The surveys use five-point Likert–scale questions, wherein students self-rated their understanding, skills, and attitudes for various topics that were covered in the course (Supp Info 6 provides sample questions). Answers ranged from "A Great Deal" to "Not at All," with a "Not Applicable" option also available. To facilitate calculation of scores, possible answers were given numerical values as follows: "Not Applicable" = 1, "Not at All" = 2, "Just a Little" = 3, "Somewhat" = 4, "A Lot" = 5, and "A Great Deal" = 6. The surveys also include free-response questions, enabling participants to offer suggestions for course improvement.

A chief goal of our assessments was to determine if students were learning—and felt that they were learning—the concepts we hoped to teach. As a representative example, we gave surveys at the start of the Fall 2011 term, at the end of Fall 2011, and at the end of the Spring 2012 term; these results are denoted 'pre-term-1', 'post-term-1', and 'post-term-2', respectively, and are shown in Fig 6. (There are generally no new students in our Spring terms, as completion of the Fall course is required, so a 'pre-term-2' pre-survey is unnecessary.) In the pre-term-1 survey, students rated their understanding of the conceptual topics we covered at a mean value of 3.62 (SD=1.16); the average was 4.78 (SD=0.84) in the

post-term-1 survey, and 5.22 (SD=0.76) in the post-term-2 survey. When asked about their understanding of the presented topics in pre-term-1, only 23% of the students answered positively (Fig 6, left-most bar), while the means rose to 60% for post-term-1 and 84% for the post-term-2 survey. This represents an increase of 61% from the start of the course (i.e., pre-term-1). The students self-assessed their lab/research skills at a mean of 4.48 (SD=1.00) in the pre-term-1 survey, 5.00 (SD=0.77) in the post-term-1 survey and 5.33 (SD=0.68) in the post-term-2 survey. Half of the students rated their laboratory skills positively ("A Lot" or "A Great Deal") in the pre-term-1 survey, 71% in post-term-1, and 90% in the post-term-2 survey, giving an increase of 40% from the start of the course (Fig 6, middle bars). With respect to their attitudes and enthusiasm for the subject of biochemistry (Fig 6, right bars), students reported an average of 4.65 (SD=1.06) in pre-term-1; 4.73 (SD=1.02) in the post-term-1 survey and 5.00 (SD=0.95) in the post-term-2 survey. Sixty percent of students reported positive attitudes in pre-term-1, 57% in post-term-1, and 77% in post-term-2 (an increase of 17% from pre-term-1); though there was an increase in the averages for this category, the magnitude was significantly less than in the other two categories.

These SALG data can be separated into the learning gain categories used in our rubrics (i.e., *Aims & Concepts*, *Experimental Design*, *Data Processing*, and *Broader Context*). The SALG data reveal that students rate their abilities in *Aims & Concepts* increasingly positively throughout the course: 28% in pre-term-1, 64% in post-term-1, and 91% in the post-term-2 surveys. In the *Experimental Design* category, 26% of students reported positive ratings in the pre-term-1 survey, 41% in the post-term-1, and 85% in the post-term-2 surveys. Similarly, *Data Processing* demonstrated an upward trend, with students self-reporting positive ratings of 29%, 77% and 93% in pre-term-1, post-term-1 and post-term-2, respectively. Finally, 27% of students positively rated their grasp of *Broader Context* aim in the pre-term-1 survey, 53% in the post-term-1, and 83% in the post-term-2. Overall, the fraction of students who consider BioLEd as having improved their skills and knowledge in biochemistry increased throughout the year-long course.

### Student Experiences: A Retrospective Survey

A major aim in BioLEd's development and implementation has been to teach undergraduate biochemistry majors how to conduct scientific research in a realistic setting. Were this achieved, a direct consequence should be a sustained increase in student confidence levels in their scientific knowledge and abilities, as well as a positive overall experience. A post-course survey was created (using QuestionPro) and was emailed to students who had completed both semesters of BioLEd; sample survey questions are given in Supp Info 7. The survey was conducted anonymously, and a monetary lottery was used to incentivize participation. The survey largely used four-point Likert–scale questions, ranging from "Strongly Agree" to "Strongly Disagree," and also included free-response questions.

Of the 128 students initially contacted, 56 completed the survey. Of these, 92% reported that they had earned a cumulative grade of ‘B’ or better in the two semesters; this is consistent with the typical average course grade over two semesters (an 84.6%, SD=6.75). Though some participants did not complete the entire survey, those portions that were completed were factored into the statistics for individual questions; surveys with an incomplete state were assumed to be due to testing fatigue rather than inaccurate answers. These retrospective surveys are summarized in Fig 7, and some of the findings are described in the remainder of this section.

The survey primarily aimed to address two questions: (i) did the course (or specific parts of the course) increase student confidence in their research, and (ii) did students feel that the course (or specific parts of the course) provided a deeper knowledge of biochemistry? Sixty percent of the students reported that they “Agree” or “Strongly Agree” that poster presentations gave them more confidence in their research and gave them a deeper understanding of biochemistry (Fig 7A). Seventy-two percent reported an increase in overall biochemical confidence, and 75% attested to a deeper understanding of biochemistry as a result of their collaborative manuscript writing (Fig 7A, B). Similarly, 71% of participants reported that the group meetings gave them constructive feedback to improve their research, and 67% felt that they had a deeper understanding of biochemistry because of these GM presentations (Fig 7A, B).

Many of the above aspects of the course evaluation reflect *group work*, which measures the ability of an individual to cohesively work together with others to generate a final product. Though group work can be difficult for students to manage, many reported it as a positive experience (Fig 7C): 85% of students testified to learning how to better communicate with their group members and to work with them professionally, and 78% of the participants reported that they learned how to better delegate tasks within their group. Overall, 76% reported a deeper understanding of biochemistry because of the group work inherent to BioLEd (Fig 7C, right-most bars).

Recent student participants were also asked to rate the BioLEd-based course in relation to other lab courses that they had taken (Fig 7D). Students overwhelmingly “Strongly Agreed” or “Agreed” that the BioLEd course (i) encouraged more independent thinking (97%); (ii) taught better time-management skills (87%); (iii) taught more effective scientific communication skills (88%); (iv) better prepared them to present scientific information (88%); and (v) encouraged greater confidence in their scientific knowledge (78%), versus other laboratory courses completed during their undergraduate studies.

The above results substantiate BioLEd’s goals, design, and implementation, at least in terms of the confidence levels and deeper understanding of biochemistry that students can achieve by being taught via activities that typify research environments—group meetings, compiling research results into manuscripts, collaborating on poster presentations, and so on. These elements of the BioLEd curriculum ap-pear to be vital in developing the communication and critical thinking skills necessary in science. The survey that we administered was rather thorough in order to allow detailed assessment of student learning and hints for future course refinements. The level of detail, however, possibly resulted in testing fatigue; also, a four-point Likert system can be too coarse (e.g., how 'strongly' a participant may agree/disagree with a statement varies somewhat, and is not readily controlled for). Future assessment efforts may consider finer (5- or 6-point) scales, and perhaps dividing the one monolithic survey into two discrete components.

### Assessing Content Gain via Pre- and Post-course Tests: A Vignette

A 20-question 'concept inventory' test was administered at the start of the first term, at the end of the first term, and at the end of the second term. These time-points in a year-long curriculum are labeled 'pre-term-1', 'post-term-1', and 'post-term-2', respectively (as above for Fig 6). As incentive, students received five points extra-credit for completing these quizzes. The questions were designed to address our learning gains (described above), and varied in complexity from highly practical (e.g. read the volume delivered from a pipette image) to the higher-level skills required to critically interpret kinetics data results; a control question was used that concerned material not included in the course. As shown in Fig 8, students demonstrated substantial learning gains over the year-long course. The class mean improved from 52% to 77% to 79% (Fig 8) of the questions being answered correctly, with a concomitant decrease in the standard deviation (4.49, 3.02 and 2.65 for pre-term-1, post-term-1 and post-term-2, respectively).

### Summary of Assessment Findings

Our initial assessment and evaluation of BioLEd indicates that this inquiry-driven curriculum provides a sound education in biochemical research, and that student learning is sustained throughout a full year. Students excel in each learning gain over time, as measured both objectively (pre-/post-course tests) and more subjectively (student SALGs). In addition to the assessments, feedback and anecdotal comments via the UVa course evaluation system have led to many curricular improvements. Past students have recognized the benefits of this type of curriculum, having rated the BioLEd course as more beneficial than any other lab course they have taken. Future work could include identifying sets of comparison groups for more thorough and systematic assessments of the BioLEd curriculum; for instance, control groups could be utilized, both at other institutions and as implemented at UVa (e.g., in a parallel lab section taught using a more traditional format).

## Dissemination

Developing, updating, and maintaining the instructional material for inquiry-based courses is necessarily more time-consuming than for other types of courses. For instance, laboratory manuals and bioinformat-ic questions must be updated each year to reflect frequent changes in electronic resources (databases change, merge with others, etc. [62]). Also, to improve the curriculum's content and student experience, constructive feedback from students and TAs is taken into account at the end of each term. These modifications occur both in our in-house laboratory manual (heavily relied on in the first semester) and in the general instructional materials that we develop (both semesters).

To disseminate BioLEd materials to both students and faculty/staff, a publically accessible resource is available at http://biochemlab.org. This website features portals for Instructors, Students, Proteins of Interest, and Collaborators (Fig 2). The Instructors portal offers three resources: Instruction Modules, Spectrophotometric Assays, and Assessment & Evaluation Tools. (This region of the site is password-protected; login credentials are available upon request.) The Instruction Modules section contains the eleven modules listed in Table I, each of which provides educational materials such as lecture slides, videos, readings, sample quizzes, grading rubrics (Supp Info 5), excerpts from our in-house laboratory manual, and additional resources (often from the primary literature). Through the Assessment & Evaluation portal, users can access the various assessment tools we distribute to students, as well as the results of those assessments; past assessments are also available, annotated with commentary to describe changes made to the curriculum based on the assessments. These resources are intended to assist current BioLEd instructors (at UVa) as well as external faculty/staff who wish to implement some (or all) of the BioLEd lab curriculum at their own institution. All BioLEd materials are freely available either via the website or upon request.

## Conclusions

The biochemistry lab curriculum at UVa has been revamped to provide students with an authentic research experience. Because this lab course is required for chemistry majors specializing in biochemistry, and because over 70% of our 100+ chemistry degree recipients specialize in biochemistry each year, the revamped curriculum must be scalable to large numbers of students. (The fraction of students focusing in biochemistry has steadily climbed in recent years, and may well continue to do so.) With the newly developed BioLEd curriculum described here, a vast majority of UVa's new BS Chemistry graduates will have had a genuine research experience before graduating. Perhaps most importantly, the experience that students gain in a curriculum such as this is deeply relevant to the 'real-world' situations they will face after graduation, such as the need to work effectively in a group of individuals, towards a common goal, and without a detailed protocol or rubric. The lessons that students learn in a BioLEd-like curriculum are general and transferable: whether they pursue graduate school, medical school, volunteer work, industry, or another calling, students can draw upon the resourcefulness and skills that they developed when learning how to search the primary literature for relevant information, effectively utilize web servers and other computational tools, logically design experiments, quantitatively analyze data and interpret results, and present their findings in a broader context and to a large audience of peers.

Importantly, we note that the research experiences gained in the BioLEd curriculum do not come at the expense of 'traditional' learning: Pre- and post-course tests, as well as participant self-assessments, indicate that students are learning in our four main focus areas (Fig 1). In addition, student grades improved in nearly all areas with each successive assignment. Finally, though developing the inquiry-based BioLEd curriculum was a major undertaking, its modular design allows for facile implementation by other institutions that may be interested in adopting a research-based model for undergraduate biochemistry education. To aid this, our BioLEd website provides course materials to all students and instructors.

## Acknowledgements

This work was funded by UVa (Dept of Chemistry, and College and Graduate School of Arts & Sciences), an RCSA Cottrell Scholar Award (LC), NSF DUE-1044858 (LC and CM), and NSF CAREER awards MCB-0845668 (LC) and MCB-1350957 (CM). We thank the JCSG for providing clones for many of the POIs investigated by BioLEd students in recent years, and we thank Jennifer Doudna (UC Berkeley) for helpful discussion about a year-long biochemistry lab. We also thank the many early generations of BioLEd students, TAs and other contributors, including Sarah Elkin, Jeong Hyun Lee, Lauren Lee, Elleansar Okwei, Colin Price, and Ana Wang.

## Figure Legends

Fig 1: The four learning gains assessed as part of the BioLEd curriculum (bold font in each quadrant) encompass eight focal areas, with some level of redundancy. Asterisks denote those learning foci that are addressed most intensely in the second half of the full year-long course.

Fig 2: The BioLEd website, shown in this screenshot (A), features distinct portals (red boxes) for Students, Instructors, Collaborators, and Proteins of Interest. The Proteins tab (B; yellow arrow) opens a list of POIs, arranged by enzyme class, that have been investigated by students in current or past BioLEd labs. This resource has been built and maintained using a standards-compliant content management system (WordPress), providing a modern and easily navigable framework for the BioLEd resource.

Fig 3: Group work is a core element of BioLEd's design and implementation. This schematic shows the relationship between an individual student and their lab partner (inner shell), other pairs (middle shell), and the higher-order association of groups who work on the same POI in a lab section (outer shell); the two distinct POI groups in a typical ≈20-student lab section are indicated. Within one full term, various assignments occur at either the individual level, partner level, or POI-group level (work at the POI-group level is chiefly in the second semester).

Fig 4: Experimental biochemistry is the core of the BioLEd curriculum. To obtain pure protein samples for kinetics assays, students learn standard techniques of protein induction and over-expression, cell harvesting/lysis, chromatographic purification, etc., as illustrated by this SDS-PAGE gel (and associated caption) from one of our lab section's POIs.

Fig 5: Computational biology is integrated into BioLEd in the context of protein functional annotation. Students learn that both informatics-based and physicochemical-based methodologies can be used to investigate the biomolecular sequence/structure/function relationships underlying biochemistry and molecular biology (A). For instance, students learn methods such as sequence analysis (B) and molecular docking (C). Together, these complementary approaches can help elucidate the function of their POI.

Fig 6: SALG surveys reveal positive response rates for three criteria: understanding of biochemical concepts (left), lab/research skills (middle), and attitude/enthusiasm for biochemical research (right). In each triad, representative data are shown for pre–term-1 (light grey), post–term-1 (medium gray), and post–term-2 (dark gray); in a year-long course, these terms correspond to the start of Fall semester, end of Fall, and end of Spring, respectively. Numerical values and further details are discussed in the text.

Fig 7: Retrospective surveys of recent BioLEd students show improvements in student scientific confidence (A), biochemical knowledge (B), and ability to work in a group (C); the curriculum also compares favorably to other lab courses taken by the students, based on the criteria listed in (D). Numerical details and further discussion are in the text.

Fig 8: Pre– and post–course concept inventory tests were used to assess student learning and retention of scientific content. This histogram plots data for pre–term-1 (light grey), post–term-1 (medium gray), and post–term-2 (dark gray); numerical values and further details are presented in the text.

## Tables

Table I: First-semester course modules

| Week | Module | Assignment type |
|---|---|---|
| 1a | Literature searches; electronic resources and tools (e.g., PyMOL) | Problem set |
| 1b | Basics of pipetting with the micropipette | Practical (wet-lab) |
| 2 | Critically reading the primary literature (in-lab discussion) | Problem set |
| 3 | Making biochemical buffers and solutions | Calculations |
| 4 | Enzyme kinetics (using LDH as a test system) | Lab report |
| 5 | Analyzing enzyme kinetics data (computer lab) | |
| 6 | Computational biology, I: Bioinformatic tools, web/database resources | Problem set |
| 7a | General molecular cloning and transformation | Lab report |
| 7b | Recombinant protein expression and SDS-PAGE | |
| 8 | Protein purification, I: Gel-filtration and ion-exchange chromatography | |
| 9 | Protein purification, II: Affinity chromatography | |
| 10 | Quantitative protein concentration determination; ligand-binding | Lab report |
| 11 | Computational biology, II: Molecular visualization, modeling, docking | POI report |

Table II: Second-semester timeline

| Week | Module | Assignment type |
|---|---|---|
| 1 | Organize reagents and buffers and finalize protocol for assay | — |
| 2-3 | Express, purify, and quantify POI | Revised POI report; chemical order request (reagent inventory) |
| 4-5 | Find workable solution/buffer conditions, optimize protein concentration for enzyme kinetics assays, establish controls for these assays | — |
| 6 | Group meeting preparation; evaluation of assay; begin determining kinetics parameters ($K_M$, $k_{cat}$, etc.) | Draft of *Materials & Methods* section of final POI report |
| 7-8 | Experimental determination of kinetic parameters for each POI | Group meeting presentation 1 |
| 9 | Present group meeting 1; begin systematic variation/perturbations of kinetics assays (e.g., substrate variation) | Drafts of *Introduction*, figures, and *Future Directions* sections |
| 10-12 | Repeat assays for statistical replicates or troubleshooting; final refinements and kinetics data-collection | Group meeting presentation 2; poster presentation; final POI report (manuscript format) |

**Fig 1**: The four learning gains assessed as part of the BioLEd curriculum.

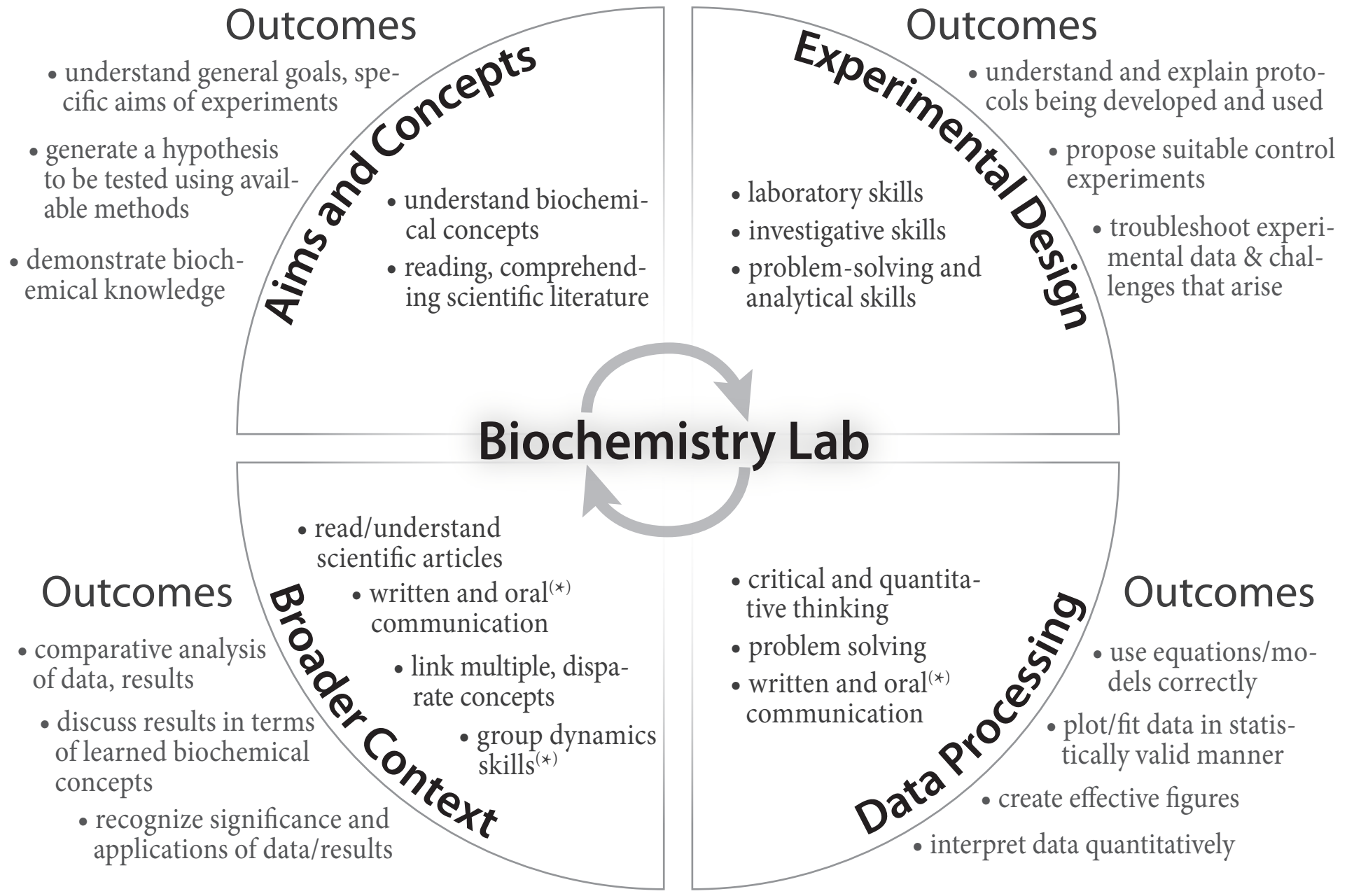

**Fig 2**: The BioLEd website, shown in this screenshot, features distinct portals (red boxes) for Students, Instructors, Collaborators, and Proteins of Interest.

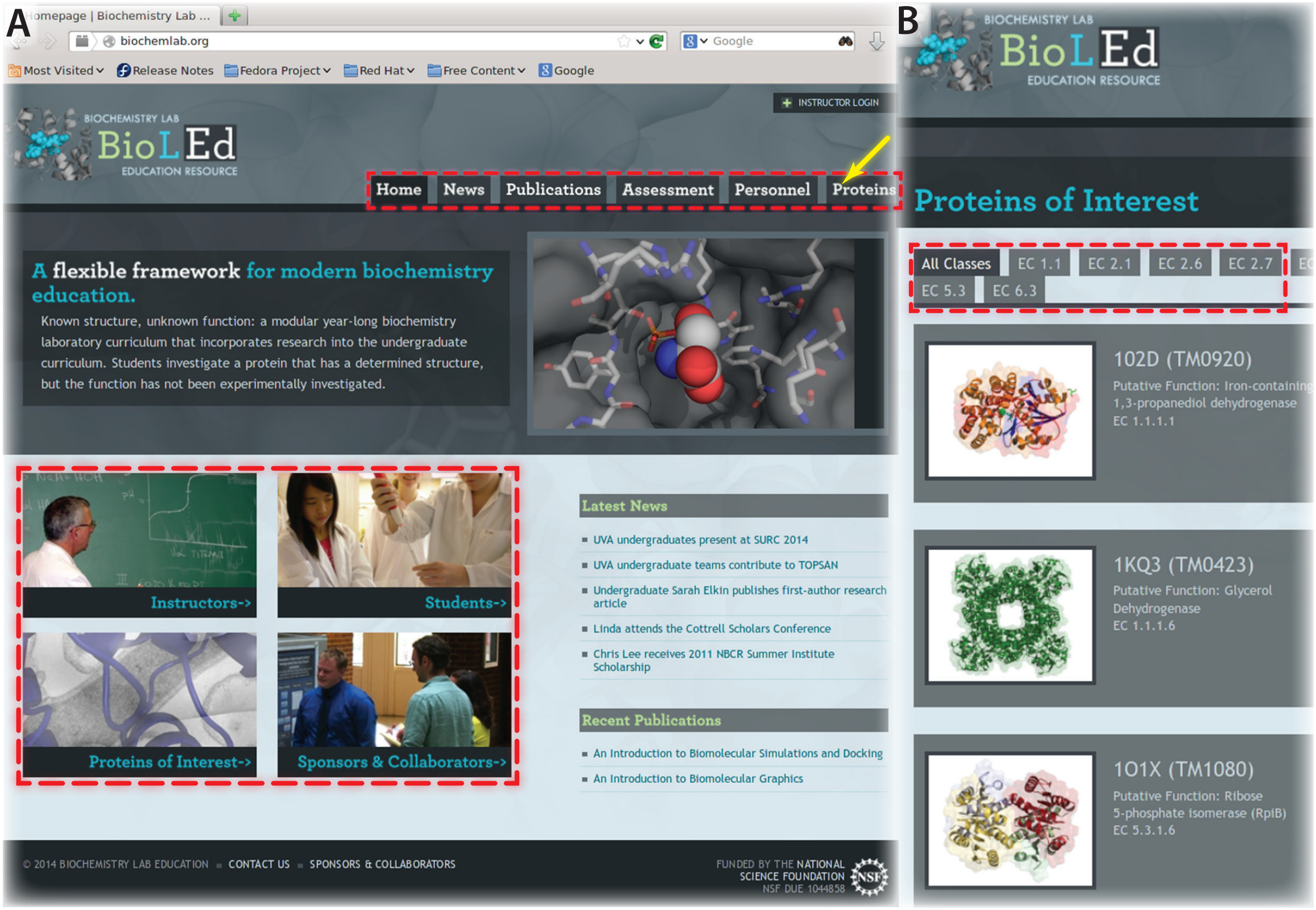

**Fig 3**: Group work is a core element of BioLEd's design and implementation.

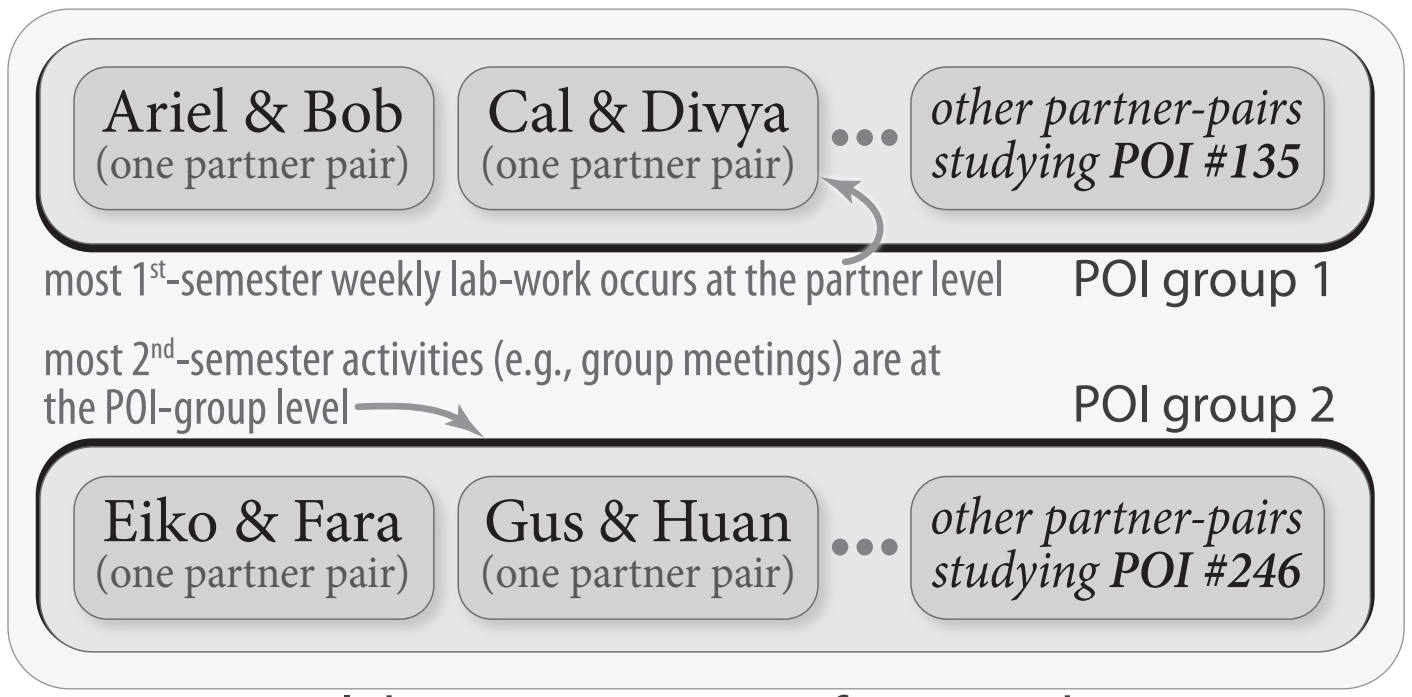

one laboratory section of ≈20 students

**Fig 4**: Experimental biochemistry is the core of the BioLEd curriculum, as illustrated by this SDS-PAGE gel of the purification progress of one of our lab section's POIs.

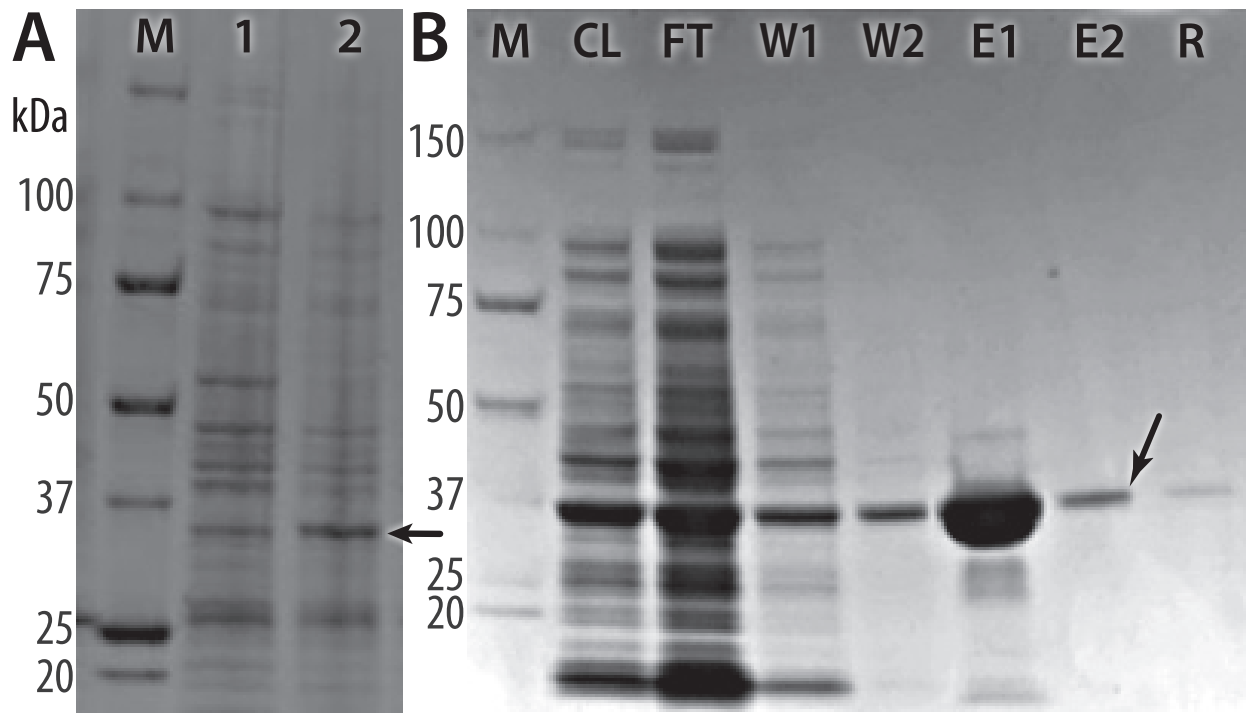

Fig N: Purification of TM0378. (A) Induction and expression: TM0378 was cloned under the control of the $P_{BAD}$ promoter, and its expression was activated and induced by arabinose. Lane M: BioRad Precision Plus$^{TM}$ Unstained Standards. Lane 1: Pre-induction sample of transformed *E. coli* culture. Lane 2: Post-induction sample. The arrow indicates successful induction of TM0378 as represented by the dark band near the 37 kDa marker. This band is consistent with TM0378's recombinant monomer MW of ≈37 kDa as calculated in ExPASy. (B) IMAC purification of TM0378: Lanes: M, Kaleidoscope MW Standards; CL, cell lysate sample prior to IMAC chromatography; FT, fraction corresponding to the flow-through fraction; W1 and W2, fraction corresponding to the first and second wash fractions, respectively; E1 and E2, the first and second elution fractions; R, resin. The arrow indicates recombinant monomer TM0378 (≈37 kDa).

**Fig 5**: Computational biology is integrated into BioLEd in the context of protein functional annotation.

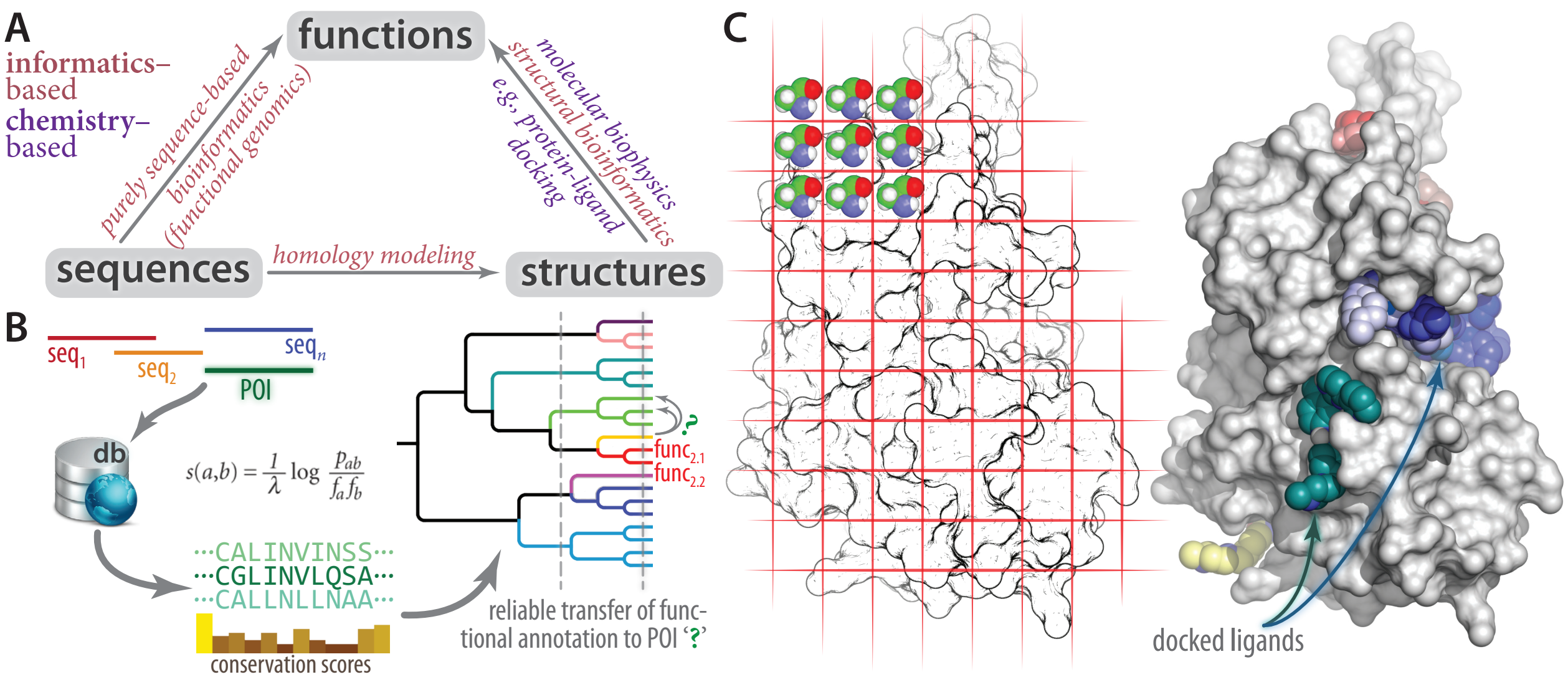

**Fig 6**: SALG surveys reveal positive response rates for three criteria.

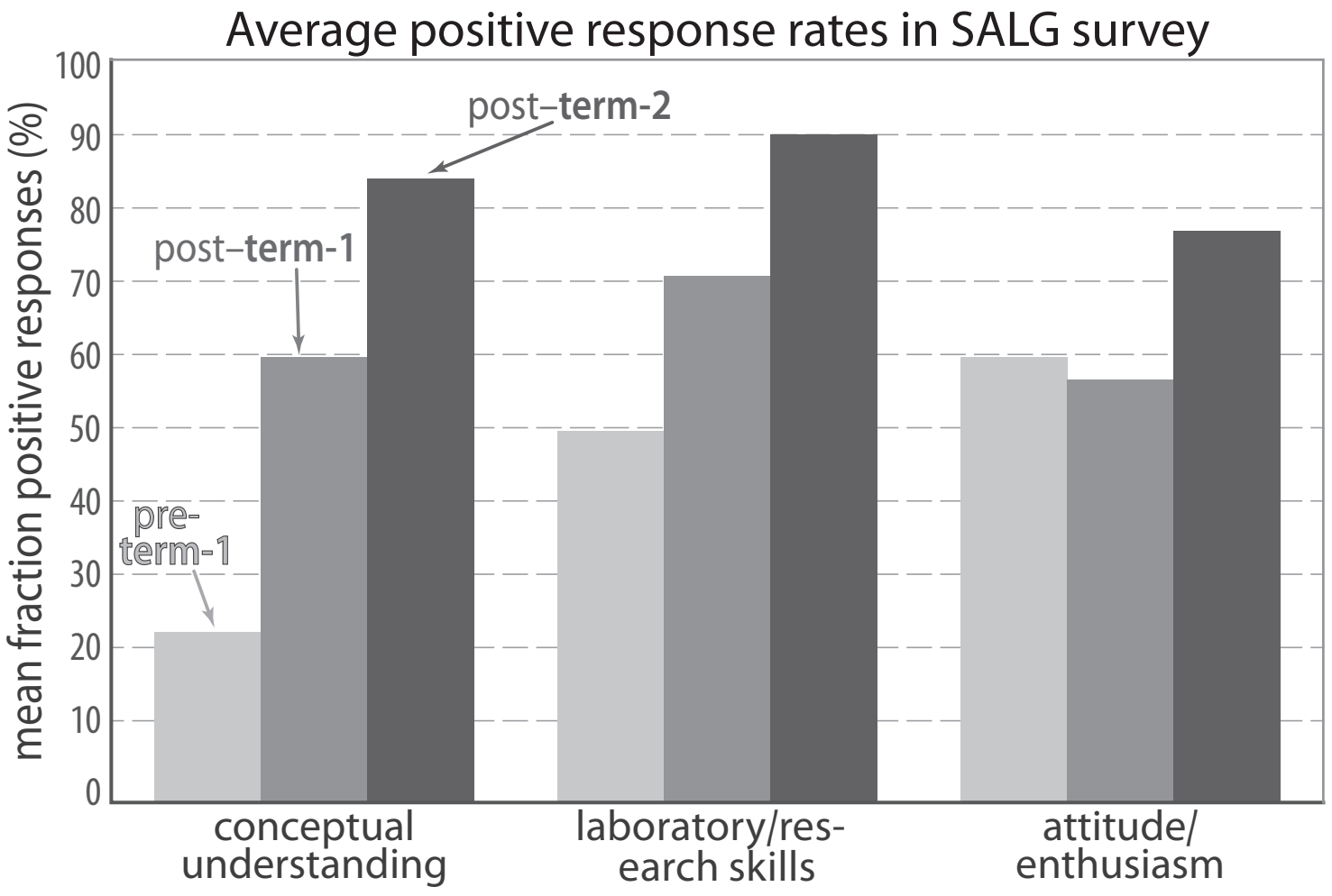

**Fig 7**: Retrospective surveys of recent BioLEd students.

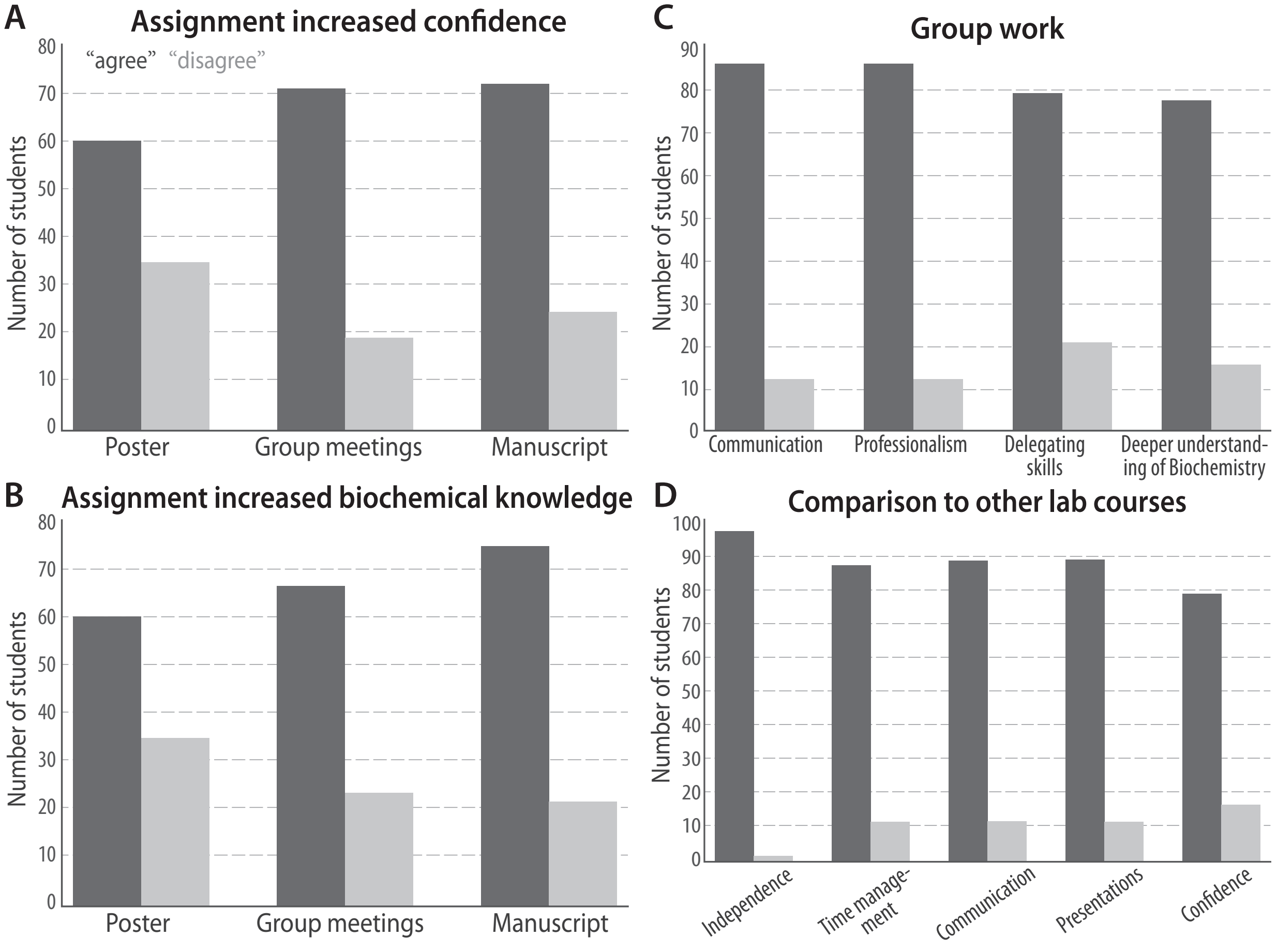

**Fig 8**: Pre– and post–course concept inventory tests.

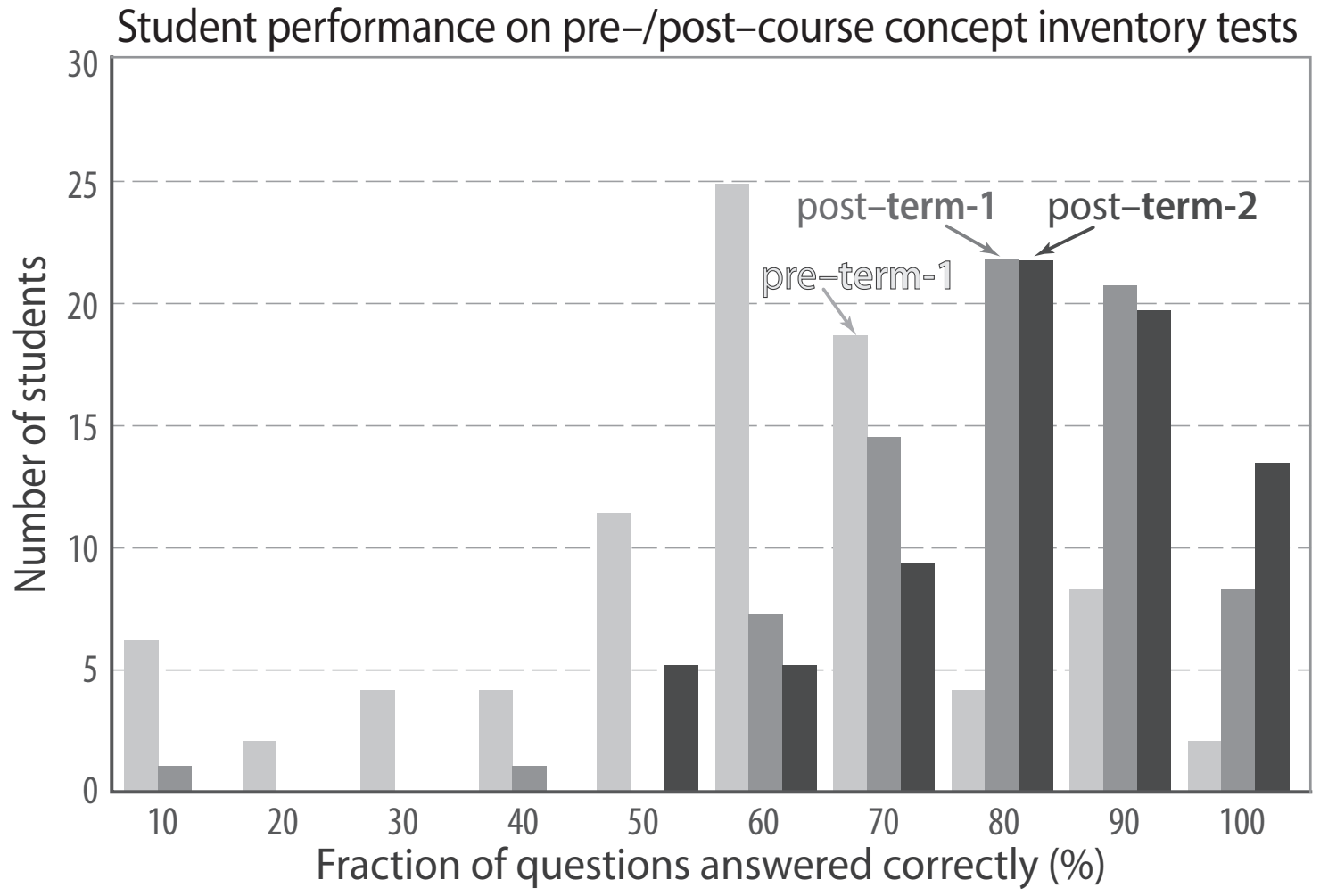

# Known Structure, Unknown Function:
## An Inquiry-based Undergraduate Biochemistry Lab Course

Cynthia Gray, Carol W. Price, Christopher T. Lee, Alison H. Dewald, Matthew A. Cline,
Charles E. McAnany, Linda Columbus, Cameron Mura

## Supplementary Information: Overview of Contents

# Known Structure, Unknown Function:
## An Inquiry-based Undergraduate Biochemistry Lab Course

Cynthia Gray, Carol W. Price, Christopher T. Lee, Alison H. Dewald, Matthew A. Cline, Charles E. McAnany, Linda Columbus, Cameron Mura

## **Supplementary Information, 1**: **Precise learning gains, organized by course modules** (see also Table I)

For each module in Table I (main text), we include here an outline of the learning gains that we expect for a student. Each assignment is graded in accord with these gains; that is, the questions or graded portions of any assignment are grouped according to learning gains (Fig 1, main text), such that we can use the assignment to help assess the learning gains for a particular topic or concept.

### Module 1A: Literature searches; electronic resources and tools

This module includes a one-hour lecture that highlights the many resources and literature search engines used by biomolecular scientists. In addition, the UVa library offers a program wherein a subject librarian is invited for subject-based instruction on current assignments and projects, with a focus on article databases, plagiarism, and critically evaluating resources. The learning gains assessed are (i) overall understanding of the purpose of literature searches (aims and concepts), (ii) identification of relevant literature (broader context), and (iii) the ability to find and understand resources (aims and concepts).

### Module 1B: Basics of pipetting with the micropipette

Students deliver water to a weigh-boat on a balance and determine the volume delivered (based on the measured mass and the density of water) from each of their pipets (a P1000, P200, and P20). They calculate standard deviations for each measurement. The learning gains assessed for this assignment are (i) pipetting skills (aims and concepts) and (ii) quantitative skills (data processing).

### Module 2: Critically reading the primary literature

This module includes both an interactive instructional period and a follow-up assignment for the laboratory period. The goal of this module is to (i) define what the primary literature encompasses (peer review, publication frequency, citations); (ii) address the differences between the *primary* (a direct report of research results and findings), *secondary* (review articles), and *tertiary* (textbooks and references) literature sources; (iii) introduce students to active and critical reading (identify the main question being addressed in the study, the conclusions, critically evaluate the data used to support the conclusions, pinpoint any missing factors or limitations); and (iv) walk students through relevant papers (i.e., provide a guided journal-reading experience). The last two goals are achieved using the C.R.E.A.T.E. method developed by Hoskins *et al*. (ref [16] in the main text). The learning gains assessed in this module are for the student to (i) understand how the data in published articles were generated (aims and concepts), (ii) critically analyze the results of each figure in the article (aims and concepts, and data processing), (iii) elucidate a hypothesis based on the results of each data figure (experimental design), and (iv) be able to propose a follow-up experiment (experimental design).

### <u>Module 3</u>: Biochemical buffers and solutions

This Module focuses on practical aspects of buffer preparation and introduces concepts of relevance to protein solutions (e.g., factors influencing solubility and stability, Hofmeister series, etc.). Students were taught the principles of making a buffer, using the Henderson-Hasselbalch equation, in their chemistry and biochemistry lecture courses. In this module, the *practical* considerations for buffer selection and solution preparation are emphasized—ionic strength, buffering capacity, compounds that may act as potential interferents in the reaction/assay, etc. Students are expected to (i) choose an appropriate buffer for downstream purification and enzymatic assays for their POI, (ii) calculate how to prepare the buffer, (iii) determine a reasonable volume to prepare, and then (iv) prepare the solution. The primary learning gain addressed in this module is laboratory skills. When students revisit these concepts in Module 9 (dialysis), the choice of buffer and how it is made can be assessed in terms of experimental design.

### <u>Modules 4–5</u>: Enzyme kinetics assay (a hands-on assay using lactate dehydrogenase or a similarly well-characterized, commercially available enzyme)

This module seeks to provide students experience with (i) performing enzyme kinetics assays, (ii) the techniques used in performing spectrophotometric enzyme assays, (iii) the process of experimental design, (iv) how to process raw data, and (v) how to analyze/interpret the resultant processed data. Students are expected to perform an extensive pre-lab assignment which requires them to think deeply about the experiment before coming to lab (the learning gain here is experimental design). While a four-hour lab session suffices for performing the experiments, completing this lab requires students to arrive well-prepared. Because of their *active* role in planning and executing the lab work in this module, students develop an appreciation of the practical considerations of experimental design and preparedness, and are more likely to carefully scrutinize protocols and attempts to make improvements ('shortcuts'), versus students who simply execute a pre-prepared protocol. In addition, students gain familiarity with each step and are thus better prepared to make logical choices when troubleshooting or adapting the protocol.

In the second kinetics module (Module 5), students learn how to analyze data from kinetics assays. Using a chosen (fixed) enzyme concentrations, students perform a series of assays in which substrate concentration is varied. They convert raw data (absorbance values) to concentrations using the appropriate extinction coefficients, generate plot(s), and calculate all possible enzyme kinetics parameters (learning gain: data processing). This module includes discussion of various means of data presentation and analysis (Michaelis-Menten, Lineweaver-Burke, Hanes-Woolf plots), determination of kinetics parameters ($K_M$, $k_{cat}$, $v_{max}$), interpretation of data from inhibitor assays, and analysis of alternate substrates. We have found that special attention must be paid to (i) how to use a spreadsheet effectively (most students use Microsoft Excel, some use Origin; we do not enforce a specific program), (ii) careful calculations of concentrations, and (iii) dimensional analysis. In working-up the data, students also learn how to effectively represent quantitative data as figures, which is a skill they use extensively in the second term. When writing the lab report for these two Modules, students are expected to demonstrate their understanding of the aims of the experiment and to relate their work to a broader context. Thus, all four of our learning gains are assessed in the laboratory and assignment associated with Modules 4-5.

Module 6: **Computational biology, I**: Bioinformatic tools, web/database resources

This module introduces students to computational methods that are commonly used in modern biochemical research. For the lab portion of this module, we draw upon an extensive and up-to-date collection of 'Education Articles' published in *PLoS Computational Biology*, including a practical tutorial on using many different types of bioinformatic approaches to analyze protein function from 3D structure (see main text). This Module's lecture materials touch upon the core ideas of (i) molecular evolution and phylogeny (including phylogenetic trees); (ii) sequence alignment methods (pairwise and multiple, substitution matrices, gaps, local/global alignment, *E*-scores); (iii) the basic idea of 'profiles' and functional annotation; and (iv) structural bioinformatics (pairwise structural alignment, finding evolutionarily-conserved functional 'patches', etc.). The last portion—3D structural analysis and an introduction to the PyMOL molecular visualization environment—supplies a natural bridge to the next computational section (Module 11).

Students perform extensive, in-depth bioinformatic analyses of their POI during lab time, and complete an assignment that details their findings. The learning gains assessed in this module include understanding aims and concepts, investigative skills, critical thinking, and broader context.

**Modules 7–9: Recombinant protein expression, chromatography, protein purification, SDS-PAGE, and dialysis**

This module introduces students to experimental techniques that are central to biochemical research, including two approaches deemed by the American Chemical Society (ACS) to be important general techniques: electrophoretic methodologies and chromatographic separations. This Module includes interactive lectures that cover (i) general methods for cloning recombinant proteins (so they learn how their POI plasmid was created), (ii) regulation of protein expression and induction in various plasmid vectors, (iii) the usage of chemical tags, such as $(His)_6$, for purification purposes, (iv) gel-filtration, ion-exchange, and affinity chromatography, (v) electrophoretic gel separation techniques, and (vi) dialysis. We have found that supplying student groups with novel, uncharacterized proteins, which they first research via the literature and bioinformatic methods, gives students a sense of ownership of the project and instills the excitement for discovery that only true research can bestow.

Modules 7–9 span four weeks, but are contained within one lab report. This lab report allows us to assess the students in each of the four main learning gains we have identified (Fig 1). Student lab reports should (i) demonstrate an understanding of the purpose of the experiments (aims & concepts), (ii) display a grasp of the methods (experimental design), (iii) feature clear figures of carefully processed data (data processing), and (iv) relate their work to the ultimate goal of characterizing their POI (broader context).

## Module 10: Protein concentration determination; ligand-binding assays

In the BioLEd curriculum, students are taught how to quantify proteins by two methods: (i) UV/vis spectroscopy (absorbance at 280nm, $A_{280}$) and (ii) a modified Bradford assay that depends on Coomassie blue binding (a BioRad assay). Also, the molecular basis of protein•ligand binding are introduced in lecture slides, and methods for analyzing such data are introduced and summarized (include equilibrium dialysis, ligand-blotting, filter-binding analysis, isothermal titration calorimetry, mobility shift assays for nucleic acid-binding proteins, and spectroscopic (notably fluorescence) measurements). In past labs, students have studied the binding of the ligand Coomassie blue to bovine serum albumin (as described in ref [18] in the main text). As for the other two lab reports in the first semester, the grading rubric for this Module 10 report also addresses the four learning gains: understanding aims and concepts, experimental design, data processing, and broader context.

## Module 11: Computational biology, II: Molecular visualization, modeling, docking

This Module introduces the basic concepts of molecular visualization and graphics (e.g., stereoscopic viewing, different types of molecular representations, surfaces, 'scenes', ray-tracing, etc.), followed by an overview of some elementary ideas of molecular modeling (e.g., rotamer libraries, homology modeling). Perhaps of greatest potential utility for their POI functional studies, we introduce the students to compu-tational methods for ligand-protein docking; in the past we have employed the PATCHDOCK server, and most recently we have begun introducing students directly to the Linux-based usage of the AUTODOCK suite (see main text). Introducing this computational Module before the second semester enables students to begin immersing themselves in the (potentially foreign) computational tools and concepts; this, in turn, leads to students (i) becoming independent practitioners of the computational methods within a matter of weeks/months, and (ii) fruitfully applying this new knowledge and computational expertise to their POI over the remainder of the year-long course. Learning gains assessed in the Module 11 assignment include data processing (to generate figures that are scientifically convincing and lucid), critical thinking (analyzing the docking results), and general biochemical knowledge (to interpret the results).

# Known Structure, Unknown Function:
## An Inquiry-based Undergraduate Biochemistry Lab Course

Cynthia Gray, Carol W. Price, Christopher T. Lee, Alison H. Dewald, Matthew A. Cline, Charles E. McAnany, Linda Columbus, Cameron Mura

## **Supplementary Information, 2**:
## A sample PyMOL-based in-class activity — Molecular visualization & structural analysis of serine proteases

Useful PDB IDs and other helpful resources (e.g., Proteopedia, http://proteopedia.org):

- 2AGI: "*The leupeptin-trypsin covalent complex at 1.14 A resolution*" (2PTN, no leupeptin)
- 2CGA: "*Chymotrypsinogen A. X-ray Crystal Structure Analysis and Refinement of a New Crystal Form at 1.8 A Resolution*"
- http://www.proteopedia.org/wiki/index.php/Trypsin, http://www.proteopedia.org/wiki/index.php/Chymotrypsin, http://www.proteopedia.org/wiki/index.php/Elastase

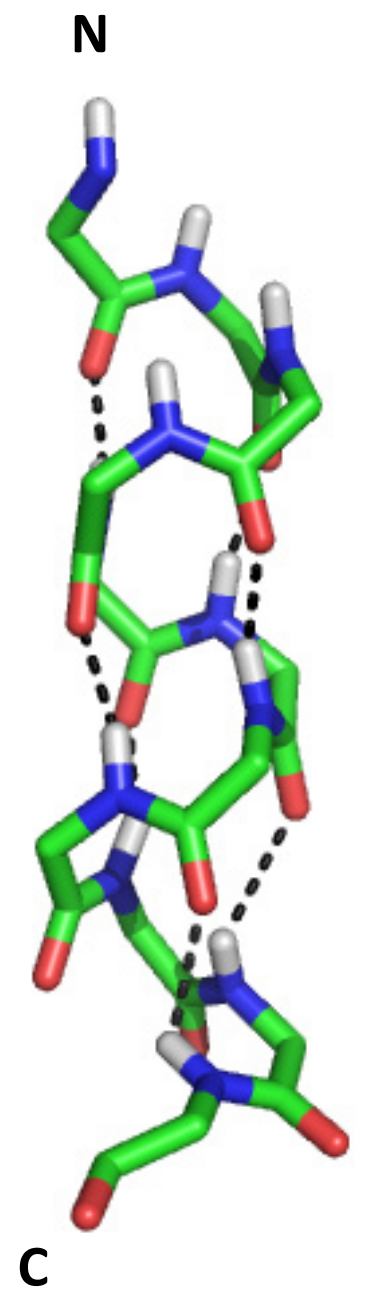

### Trypsin

- Open the trypsin PDB file in PyMOL and then:
  - Add hydrogen atoms ('A' pull-down menu → 'Hydrogens' → 'Add')
  - Issue this command: `select myhelix, resi 235:245`
  - Issue this command: `hide everything`
  - Display cartoon of `myhelix`
  - Double middle-click near the center of the helix (to center the molecule)
  - Zoom the view of the helix
  - Identify the N- and C-termini of the helix
  - Show main chain atoms as sticks (for the helix selection)
  - Set background color to white
  - Save the image, and label the termini and the H-bonds that stabilize the $\alpha$-helix

- Now, for the same helix, select the **hydrophobic residues** and color the selection (green). Hint: consider a command such as this (all one line):

  select my_hydrophobic, resn \
  leu+val+ile+gly+pro+ala+phe+met

- Now select the **polar residues** and color the selection (red), using a command similar to this:

  select my_polar, resn \
  glu+asp+asn+gln+lys+arg+his+ser+thr

  Is '`myhelix`' amphipathic? Explain.

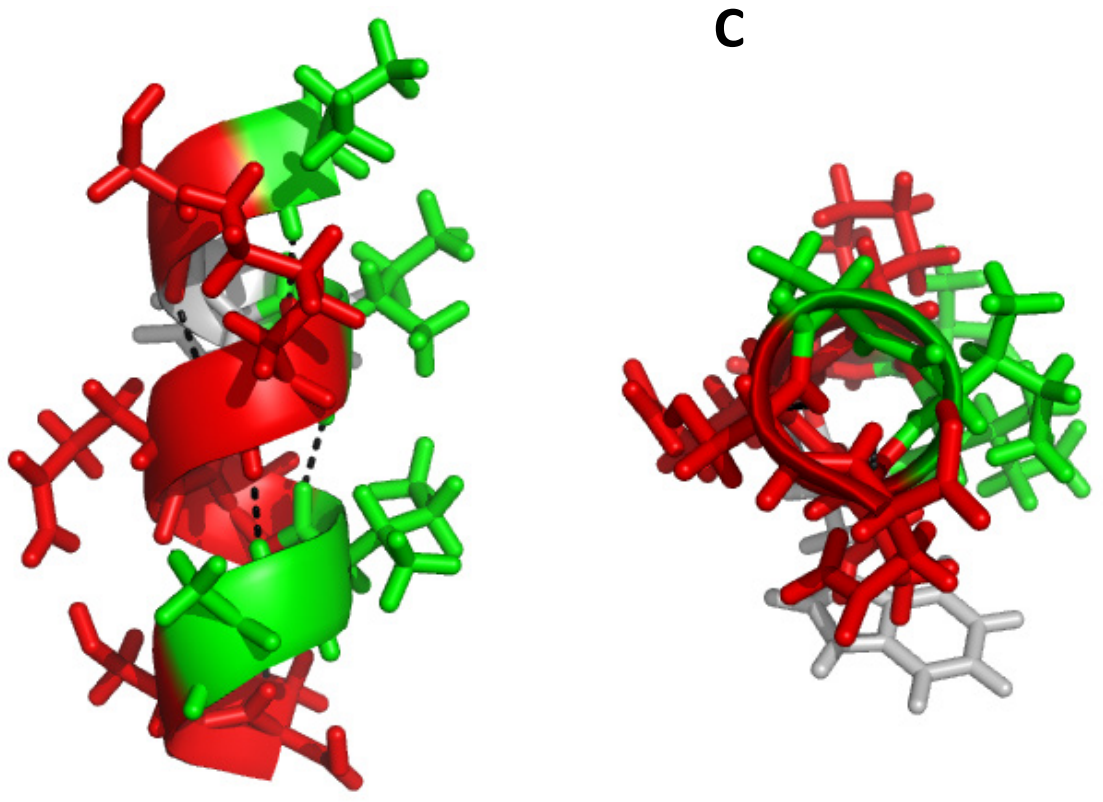

- Select residues 1-234 and color them gray.
- Now display the cartoon of the entire trypsin molecule.
- Print it out and label the hydrophobic regions of the helix and explain nature of the helix with respect to the rest of the protein.

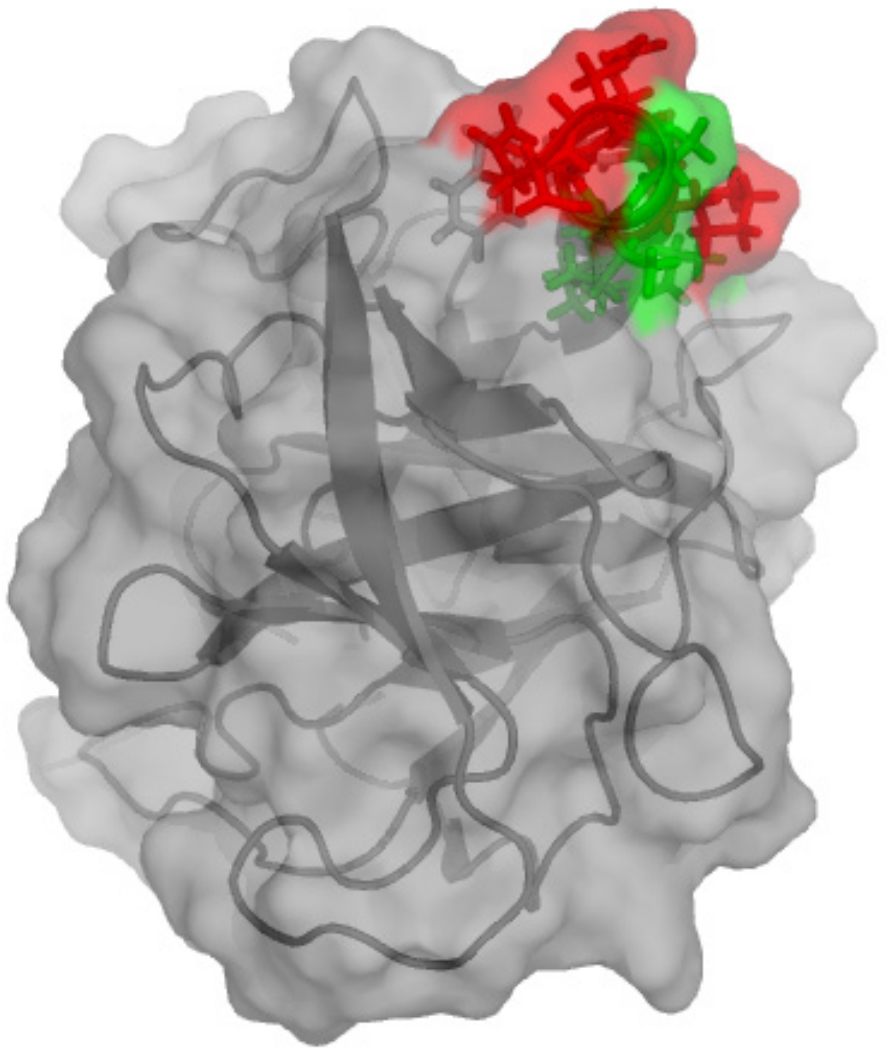

**Chymotrypsin** (chymo), another serine protease

- Open the chymo PDB file in PyMOL.
- Using the commands above, identify an amphipathic β-strand. What residues comprise the strand?

- How many disulfide bonds are there in chymotrypsin?

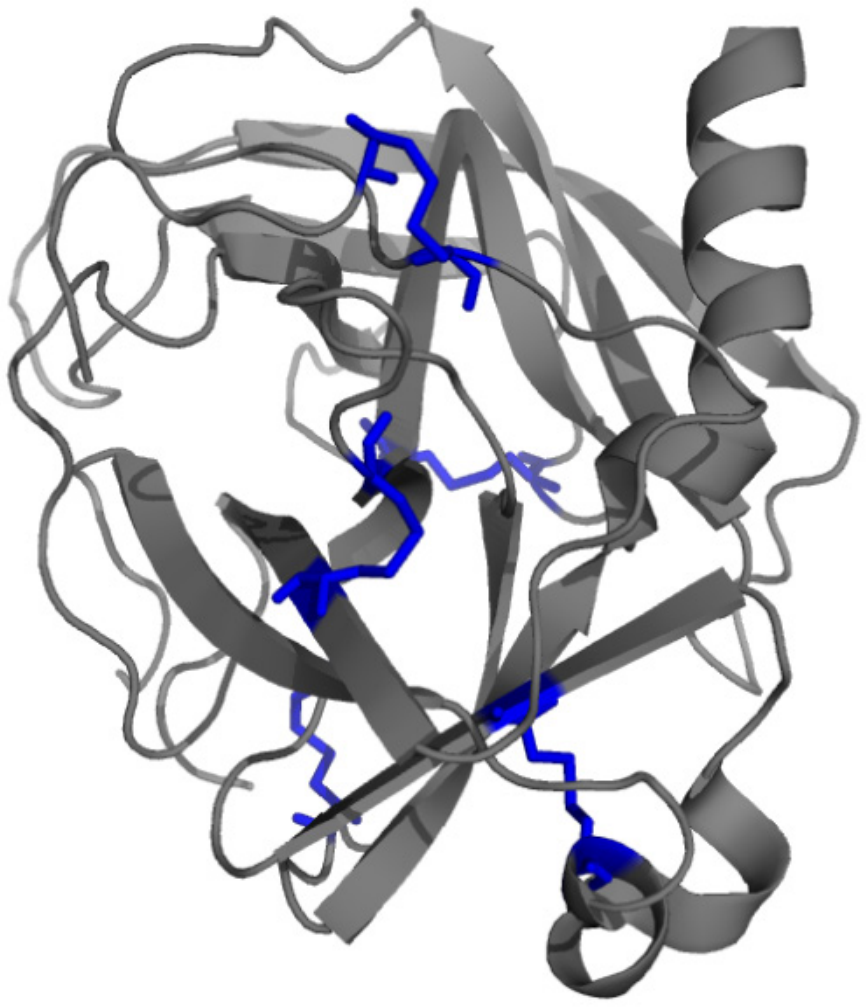

…The Protease Mechanism — substrate specificity!

- Select residues 57, 102, and 195 using this PyMOL command: ___________________________?
- Color the entire molecule gray and the selected residues another color.
- Display a molecular surface.
  *Do you see a large cavity next to the colored residues?*
- If so…
  - What properties of the cavity do you believe to be important in binding the peptide substrate?
  - Compare the cavity to that of trypsin and elastase. Do they differ? (If so, how?)
- Prepare a figure that illustrates the differences amongst these proteins (highlighting active site residues, cavities, and stabilizing residues).
- Select residues 215-219 and color them a different color (view with and without the surface).
  What is the function of these residues in the chymotrypsin mechanism?
- The compound tosyl-L-phenylalanine chloromethyl ketone (TPCK) specifically inhibits chymotrypsin by covalently labeling His57.

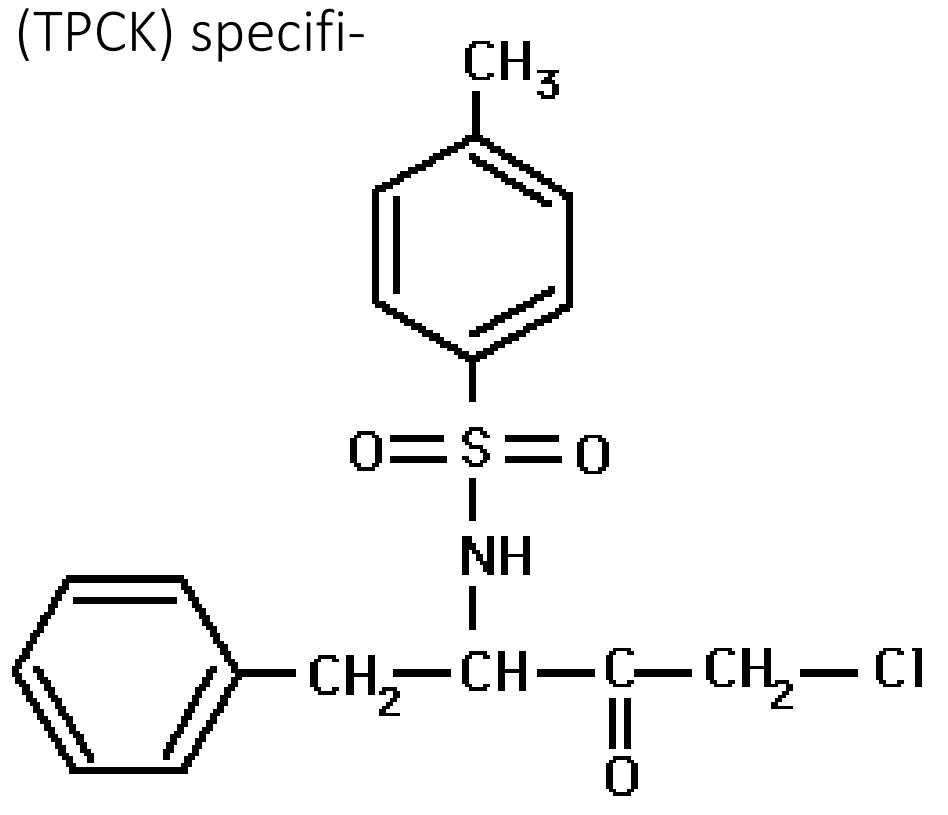

- Given the chemical structure, can you suggest a mechanism for the inactivation reaction? (You can consult the enzyme catalysis chapter in Voet & Voet or other standard *Biochemistry* texts.)

- Why is this inhibitor specific to chymotrypsin?

- Draw a derivative of TPCK that might inhibit **trypsin**, highlighting what moieties you've changed.

# **Known Structure, Unknown Function**:

An Inquiry-based Undergraduate Biochemistry Lab Course

Cynthia Gray, Carol W. Price, Christopher T. Lee, Alison H. Dewald, Matthew A. Cline, Charles E. McAnany, Linda Columbus, Cameron Mura

# **Supplementary Information 3**:

Molecular Docking Tutorial for the Biochemistry Lab (Chem4411/21)

## 1 Initial Setup, Introduction to Linux

In this lab, we will dock ligands to your POI using the AutoDock Vina software on the Linux operating system. This section contains some computing details that may seem superfluous at first, but the material is critical to the rest of the docking workflow, so please study it carefully.

### 1.1 Basic Information and Nomenclature Conventions

Here we explore the basics of using the Linux operating system, and we describe some important terminology and formatting conventions that appear throughout this tutorial. Note that new terminology is defined in context using *italics* typeface.

In the following pages, we show commands/concepts/terms in the left-hand side of the table, and matching explanations follow on the right-hand side.

| | |
|---|---|
| ***Dolphin*** | The file manager for the Linux distribution (Fedora) and window manager (KDE) that you will be using. The Dolphin system works essentially like its Windows counterpart, Windows Explorer. To start, single-click the Home icon on the Desktop. |
| ***Konsole*** | A graphical environment that places you in a Unix *shell*, which allows you to input commands as text. To open a Konsole, right click on empty space on the Desktop and select `Konsole`.* |
| ***Home Directory*** | This is the directory that is shown when you first open Dolphin. You can consider this as roughly equivalent to My Documents in Windows. It is often denoted by a '~' in file-paths. |

*The Konsole program is, technically, a *terminal emulator*, which provides you with a *command-line interface* (CLI). There are three two fundamentally distinct modes that one uses in working in Linux: (i) GUI-based (mouse clicks, like in MS Windows or Mac OS) and (ii) text-based (in the shell, using the CLI). In reality, a hybrid of (i) and (ii) is often the most efficient approach, and for this reason we introduce you to the Unix shell in this lab course. Many commands that we will use can be run only from the CLI, or can be run far more powerfully via the CLI (this may be counterintuitive, right now!). In general, running commands and performing operations in the shell will save much effort versus other methods (and is more easily reproduced, as one can communicate to someone else a list of text commands much more easily than showing mouse clicks across the graphical desktop background).

| | |
|---|---|
| `[jobDirectory]` | This is how we will refer to the directory where all of your work will be done for a single *job*. (Think of a job as one small, self-contained unit of work; for example, it would be one replicate, if you were pipetting many solutions to repeat a wet-lab experiment in triplicate... In computational biology, you would say you performed the calculation, or *job*, three times.) You will create your `[jobDirectory]` in the next step of this tutorial, and you will need to navigate to it on several occasions. |
| `text` | We will use `this formatting` to highlight many words throughout the tutorial. `This font` indicates one of two things, depending on context. First, you are looking for a button, field or file called `some_name`. The other case is that you will be typing a text command using the keyboard. In both cases, it is `text` that you should find verbatim, unless... |
| `[text]` | Text with `[square brackets around it]` will be text that is not precisely the same for every use. This will be such things as PDB codes or ligand names, which will generally differ for each job. |
| ***GUI*** | This stands for Graphical User Interface, which is how you usually interact with your computer. |
| ***PGUI*** | This is a denotation that will be used when the command is in the small gray PyMOL box containing the `File` menu. 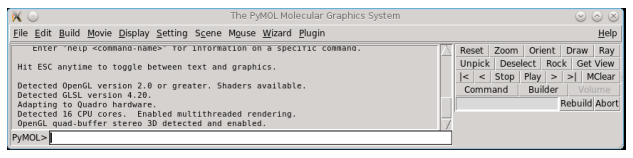 |
| ***VGUI*** | This is a denotation that will be used when the command is in the PyMOL Viewer. Most commands given here will be on the right side panel (the graphical menu of buttons). 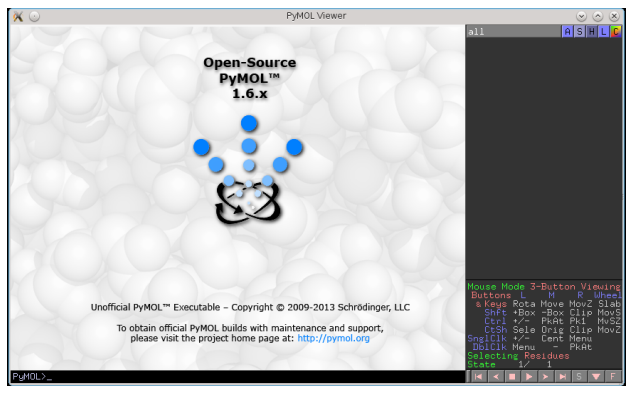 |
| ***AGUI*** | This is a denotation that will be used when the command is to be issued in the AutoDock plugin. Make sure you check the tabs at the top, if you are having a hard time finding a button. 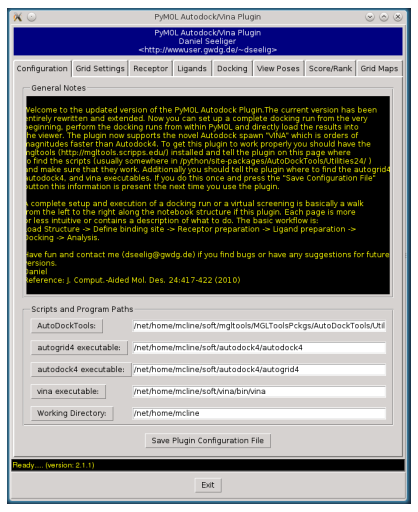 |
| PyMOL>[cmd] | This indicates that you should type [cmd] in the PyMOL shell (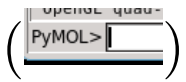). Feel free to use PyMOL's GUI for any commands that you feel more comfortable with, but note that it is often simpler for us to give precise instructions by using text commands, for reasons described in the footnote on page 1. (Also, as you learn PyMOL's text commands, you will become faster and more versatile in PyMOL.) |
| :)[cmd] | This indicates that you should type [cmd] on the Konsole command line (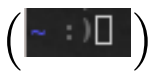). Note that :) is **not** part of the actual command, but rather it denotes the **shell prompt**; so, do not type a :), instead just type that text following immediately to the right of the closing parenthesis. |
| :)cd [dir] | This command, which stands for change directory, allows you to navigate the filesystem while in a shell (in *Konsole*). [dir] refers to the directory that you wish to go to, so to move to a directory called foo, you would type :)cd foo. Some special symbols can occur in the place of [dir] in order to do specific things. For example... |
| :)cd .. | A command that moves you up in the filesystem by one directory. So, if you are in ~/TM1689.mcline/test1, then :)cd .. would move you to ~/TM1689.mcline |
| :)cd | In this special case of cd (when no directory is specified) you will be taken back to the Home Directory. |
| :)ls -l | This extremely usful command shows a listing of all files in the current directory (analogous to seeing a list of all files graphically, in the Windows or Mac OS). |
| :)pymol | This command launches PyMOL from Konsole. Note that you can also maneuver the file-system from within PyMOL, using cd, in the same fashion as from within a shell (Konsole). |
| <TAB> | This denotes a literal TAB on the keyboard. We use this key often because <TAB> is a powerful tool when using the shell (Konsole) and from within PyMOL's command line. <TAB> triggers the computer to try and finish that which you began typing (this is known as *tab completion*). This means that if you have a long command name, for example autoligand, then, more often than not, it will suffice to begin typing aut and then press the <TAB> key and let the computer finish your thought. If there are multiple (reasonably few) options for completion of a command that begins aut..., then the shell will list those potential commands, and that is handy in its own right (e.g., when you know only the begining of a command, or can't remember what some file was called...). |

### 1.2 Setting up Your Directory

In this section, you will create your `[jobDirectory]` and start PyMOL in Linux for the first time.

1. Open the home directory from the Desktop icon `Home`
2. Right click in Dolphin.
3. `Create New` → `Folder...`
4. Name the Folder `[POInumber].[userID]`, where userID is your UVa computing ID and POInumber is the ID code for your protein of interest (POI). For example, I would use `TM1698.mac7yx`
5. Enter the folder that you just created by single-clicking on it (a single click is often enough for this Linux environment; you don't necessarily need to double-click like in Windows).
6. Repeat the above method to create a new job folder called `test1` (or whatever you would like) in `[POInumber].[userID]`. This is where a single job will be performed. If a new job is done then create a new folder in `[POInumber].[userID]` to work in.

## 2 Performing a Docking Calculation Starting with Sanitized Receptors and MOL2 ligands

Before we can begin any docking project, we must gather the necessary files into your `[jobDirectory]`.

1. Open the Home Directory and navigate to the `RECEPTORS` directory.
2. From the receptor directory, click once on the appropriate `[receptor].pdb` file and use the Ctrl-c keystroke to copy the file (alternatively, right-click on the file to see the list of possible options, one of which should be to copy it).
3. Navigate to your `[jobDirectory]`, which is the directory we named `test1`.
4. Use Ctrl-v to paste the file into your `[jobDirectory]`.
5. Now navigate to the `LIGANDS` directory in the Home Directory and repeat the copy-and-paste procedure for the `[ligand].mol2` file.
6. Open Konsole and navigate to your `[jobDirectory]`, like so: `:)cd [jobDirectory]`.
7. Open PyMOL from within this Konsole by typing `:)pymol`.

Now, we can set-up and run our docking calculation from within PyMOL using the following sequence of operations in the PGUI window:

| | |
|---|---|
| `PyMOL> load [ligand].mol2` | For the ligand, we are using the MOL2 coordinate file downloaded from the ZINC database. If you ever need coordinates for a ligand, we advise that you start searching at the ZINC database (an online database of purchasable ligands), as your search may well end there or never be complete. Note that no pre-processing has been performed on the ligand prior to your receiving it here. |
| `PyMOL> load [receptor].pdb` | We use Protein Data Bank (PDB, `.pdb`) structure files for docking. The PQR format (that was also in `RECEPTORS`) is useful in electrostatics calculations, but that can be a topic for later analysis or discussion. |
| `PyMOL> h_add [receptor]` | This command adds hydrogens based on empty valences. (As a sidenote, the PyMOL protonation tool is necessarily the 'best' algorithm, but its advantages are that it does not require other third-party programs or libraries, and it isn't as picky about ligands in the PDB file (versus other methods). |
| PGUI: `Plugin` → `Autodock/Vina` | This opens the GUI plugin that we will use to set-up our docking calculations. The GUI should open to the `Configuration` tab, which should already be set with appropriate parameters. |
| AGUI: `Grid Setting` | In this tab, we will set-up the three-dimensional (3D) region of space where the molecular docking will occur. |
| AGUI: In `Calculate Grid Center by Selection`, type `[receptor]` in the field. Press Enter. | This places the center of the calculation grid (last step) at the center-of-mass of your protein, which is a good starting point; we may end-up needing to adjust this in a moment (see below). |
| AGUI: In `Parameters`, change `Spacing` to `1.000`. | In AutoDock Vina, a 3D grid is laid over the protein, and the interaction energy of various atoms is computed at each grid point. By setting the spacing to 1.0, the other measures presented in the GUI will also be in units of Angstroms (Å), and thus more easily understood. (If this step is confusing, that is ok: the PyMOL plugin is smart enough to adjust your measurements to correct geometric amounts when it creates the configuration file.) |
| AGUI: In `Parameters`, adjust `X-points`, `Y-points` and `Z-points` until the grid box covers the protein. | In this step, you are telling AutoDock Vina where to search for the potential ligand-/substrate-binding site. For faster calculations, you will want this grid to be as small as possible. You can also adjust the `Grid Center Coordinates` to help shrink this region. |
| AGUI: In `Config File`, press `Save` | This saves the coordinates for the grid to a file called `config.txt`. IMPORTANT: This is probably the first time so far that it has become crucial to have done all of the work in **your** `[jobDirectory]`. If, instead, you had been working in the Home Directory (where multiple users where working), then only one `config.txt` would have been saved in the Home Directory (the others would be over-written), and so odds are it isn't yours! This can generate great confusion. |
| AGUI: `Receptor` | In this tab, we will finish preparing the receptor for docking by saving AutoDock's own special format, PDBQT, which stores some additional information (beyond the coordinates in the PDB file format). The most critical piece of additional data is the bonding information for all atoms in the system — this information defines the molecular *topology* and also enables us to specify which bonds we will allow to freely rotate. |
| AGUI: Select `[receptor]` from the PyMOL selections. | Here, you are simply telling PyMOL which of the objects that it is storing is the *receptor* (i.e., your POI, which is to be docked to). |
| AGUI: Press `Generate Receptor` -> | The PyMOL plugin will now go find the correct preparation script and will apply it to the receptor (your POI). So, just wait for it to finish and add your receptor to the `Receptors` list. While this is occurring, you should look in the `Log` field for any errors, because if any part of this setup was wrong then this step is likely to fail (not to worry, this is probably not your fault). Unfortunately, these error messages can be subtle and, sometimes, the program will continue on computing, but will give flawed results. If an error arises here, and if you research it a little (use Google) and do not understand it, please show your TA the error message (it may be a computer/IT problem that can be readily addressed by one of us). |
| AGUI: `Ligands` | This is exactly the same as the receptor (above), except that now you are chosing your *ligand*... So, give this a shot on your own. |
| AGUI: `Docking` | This is where we can print the final configuration file for AutoDock Vina. The `Run Vina` button seems to be broken (software is not always perfect), and so we will have to resort to the Konsole to actually run Vina. |
| AGUI: Press `Write Vina Input File(s)` | The program writes another file in your `[jobDirectory]`, which is probably starting to look like a cluttered mess. That's OK. |
| Open a new Konsole and `:)cd [jobDirectory]`. | There is a shortcut to do this, actually: In the Konsole that is running PyMOL, double-click the free area at the bottom, located beside the current tab. This opens a new tab which provides a shell that is already in the directory of the previous tab (so you don't have to navigate there again). |
| `:)vina --config [ligand].vina_config.txt` | By executing this command — type it exactly as shown, and press Enter — your computer should happily begin computing docking conformations. When this finishes, we will begin the fun part, *analysis* of the docked structures of the ligands to your POI (each of these are known as docking *poses*). Wait for this job to run to completion, which will be apparent when the command prompt `:)` returns control to you (the user) rather than the program that just finished running. |
| `:)dockProc -csv log.csv [ligand].vina.log` | This executes an in-house *script* that reads the log file and builds a simple table in comma-separated value (CSV) format; MS Excel or most other data-processing/math software can read/import such files. |
| Move `receptor.[receptor].pdb`, `[ligand].docked.pdbqt` and `log.csv` to your computer (e.g., you can email it to yourself). | Analysis of the docking results does not require the plugin nor any Vina software, so we can complete that stage of our work on any computer workstation with PyMOL installed (e.g., your laptops). Make sure that you save these results in a place that you can find. If you prefer to continue on the Linux platform, we have several workstations in the research lab that you can ask us about using. |

## 3 Analyzing Docking Results: The Mechanics

In this section, we will load the docked ligand conformations (the poses) into PyMOL for further analysis... and that will be all that is covered in this current tutorial, because analysis of the docked poses — literally, the docking results — is your job, and is specific to your POI. (Note that by 'analysis' we mean visual analysis and interpretation of the locations of the ligands [on the POI], their detailed 3D structures, inter-atomic interactions, ligands···POI contacts, etc.)

1. Open `log.csv` in Excel or a comparable program. If you are using a Linux workstation, we suggest LibreOffice Calc. Note that there are no headings. This is because we wish for this file to be easily loaded into any program that accepts CSV, but headings may hinder such compatibility. The headings are, from right to left, 'Ligand Identifier', 'Mode Number', 'Binding Affinity (kcal/mol)', 'RMSD upper bound (Å)' and 'RMSD lower bound (Å)'. The RMSD values are of limited value, particularly when you dock to the entire protein (known as *blind docking*). The binding affinity can be viewed (very roughly) as the thermodynamic binding affinity, were the ligand to bind in exactly that pose; however, these are not truly accurate $\Delta G^{\circ}_{bind}$ values, and are only particularly useful when internally compared across different docked conformations/ligands/etc.

2. Start PyMOL on your computer.

3. Use the PGUI: `File` → `Open...` to find and load in your `receptor.[receptor].pdb` and your `[ligand].docked.pdbqt`

4. The docked poses are now in one PyMOL object with multiple states. To switch between the states you can use the arrows at the right of the PyMOL Viewer.

Congratulations! Now that you're familiar with the mechanics of a docking calculation, use what you already know about PyMOL and biochemistry to draw conclusions from the docked conformations; ideally, perhaps you will be able to assess the ligand-binding preferences of your POI.

# Appendices

## PyMOL: A Quick-start Guide

### Installation

There are two major ways to install PyMOL. First, one can obtain the educational version, though that edition is some releases behind the latest production version. Nevertheless, the educational version can be installed via a relatively simple process, and if you wish to use this version, Google 'pymol' (follow the directions at http://www.pymol.org). The second method requires you to compile PyMOL from source-code; this considerably more complicated route does provide you with the very latest, 'bleeding-edge' version. For Windows, visit http://www.pymolwiki.org/index.php/Windows_Install for directions. For Apple, go to http://www.pymolwiki.org/index.php/MAC_Install. Both methods may take some tinkering and online searching in order to make sure that appropriate libraries are in-place, cross-compatibility with versions of the Python and Tcl programming languages is not a problem, and so on.

### Navigating in PyMOL

| | |
|---|---|
| ***Object*** | This organizational unit is how PyMOL internally stores a 3D structural entity. When a protein or any other molecule is opened in PyMOL, that auto-creates one *object*; the next molecule that is loaded will be a new object, and so on. These objects can be edited as one group. |
| ***Object Control Panel*** | This is the area on the right-hand side of the PyMOL Viewer providing a list of the objects. Many of the GUI commands will be found here, and we will assume that you can explore this area on your own. |
| Left-click & drag | This rotates the protein representation in 3D space. Play with this for awhile to become comfortable with how this works. |
| Right-click & up-down drag | This zooms in and out on the protein. |
| Scroll wheel | This changes the *clip*, which is the width of a slab that dictates how much depth of the 3D space (the $z$-direction) is rendered at once. Most of the time, it's not a bad idea to begin by increasing the clip until the entire protein can be seen (see also PyMOL's closely related 'zoom' and 'center' commands). Another way to acheive this is to type `zoom` in the PyMOL PGUI. |
| `PyMOL>load [file]` | This loads a structure file (e.g., in PDB format) into PyMOL, and thereby *instantiates* a new object corresponding to this structure. |
| `PyMOL>save [file], [selection]` | This is the command for all of PyMOL save functionality, so it is a bit intricate. First, you specify the `[file]`, which is what the file will be called. This needs the extension because that is how PyMOL determines in what file format to save. The two important types are `.pse`, which is a PyMOL session file allowing you to save your work, and `.pdb`, which simply specifies a 3D structure in PDB format. |
| `fetch [pdbCode]` | This automatically retrieves the PDB entry from the PDB database, without your having to explicitly download it first (in fact, on Linux the PDB file will be downloaded to the local directory from which PyMOL was launched). |
| `PyMOL>orient [object]` | This resets the view to see the `[object]`. |
| `PyMOL>delete [object]` | This removes the object from PyMOL. |

## Selections in PyMOL

Atom selections are a vital part of being able to manipulate molecules and subsets of molecules in PyMOL (or any other molecular visualization software environment). For high-quality molecular graphics, you will have to become quite familiar with *named atom selections*. Selections can be thought of as a type of object, but can contain any logical set of atoms, which can then be manipulated together as a unit (by 'logical' we mean in a Boolean sense). You can make selections with text commands or by clicking on the protein. The click method has seven modes for different selection *scopes*: atoms, residues, chains, segments, objects, molecules and C-alphas. To change the mode, PGUI: `Mouse` → `Selection Mode`.

| | |
|---|---|
| `PyMOL>select [selectionName], [descriptors]` | This is a command that makes a selection in PyMOL using logical descriptions. The `[selectionName]` is what the selection will be called in the Object Control Panel, and `[descriptors]` is the logic statement for whether or not an atom belongs in the selection. How to form the logic statements will be the rest of the topic of this section. If the `[selectionName]` is omitted, then the name will default to simply 'sele'. |
| `[object]` | When an object is included as part of the descriptor, then an atom must be part of that object in order to be chosen. So, if you would like to select all atoms in your receptor, the simple command would be `PyMOL>select sele, [receptor]`. This isn't useful in and of itself, but will often be used in logic statements (when multiple objects are loaded, e.g., your POI and a homolog to be used for structural alignment in PyMOL). |
| resn | This is a descriptor that means residue name. So, if you wish to select all atoms associated with a residue that is named 'PLP' (in the PDB file from which the object arose), then the command would be `PyMOL>select sele, resn PLP`. |
| index | This is a descriptor that means atom index number. So, if you wish to select atom 1 of the protein then the command is `PyMOL>select sele, index 1`. For early work, this descriptor is likely not as useful as others, because the mouse can achieve the same functionality (without your having to know the atomic index number(s)). |
| resi | This is a descriptor that means residue identifier. So, if you would like to select residue 1 of the protein, then `PyMOL>select sele, resi 1`. Again, this may not be as useful initially because the mouse can accomplish much the same without your needing to know the residue identifiers. However, a useful feature here is the ability to use this descriptor to select either a contiguous range of residues (e.g., 'resi 1-10') or a disconnected set of residues (e.g., 'resi 1,3,5,7') |
| symbol | This is a descriptor that means chemical symbol. So, if you seek to select all nitrogen atoms in an object, then `PyMOL>select sele, symbol N`. |
| chain | This is a descriptor that choses all atoms at the chain level. So, if you would like to select all (all atomic entities) in chain A, then issue the command `PyMOL>select sele, chain A`. This descriptor becomes a usedful part of the selection logic when dealing with oligomeric (multi-chain) objects, such as is the case with many POIs. |
| hetatm | This special descriptor symbolizes every atomic entity in the object that is not part of the protein – i.e., is not proteinaceous (e.g., water molecules, bound ions, etc.). The simplest example of a command using this descriptor is `PyMOL>select sele, hetatm`. This selection *macro* gets its name from the fact that 'hetatm' is the starting string in these non-amino acid lines in PDB files. |
| not | This operator modifies an otherwise 'normal' atom selection string by (logically) negating it. An example would be `PyMOL>select sele, not hetatm`, which would select all non–hetero-atoms (i.e., the protein). |
| and | This boolean logical operator combines two descriptors by selecting only those atomic entities that satisfy both descriptors (i.e., it is the logical *intersection*). An example would be `PyMOL>select sele, [object] and symbol C`, which would select all of the carbon atoms in `[object]`. |
| or | This boolean logical operator combines two descriptors by selecting only those atomic entities that satisfy at least one of the descriptors (i.e., it is the logical *union*). An example would be `PyMOL>select sele, symbol N or symbol O`, which would create an atom selection containing all of the oxygen and nitrogen atoms in the object. |

## Modifying the Molecular Scene/Representation

| | |
|---|---|
| `PyMOL>color [color], [selection]` | This colors the selection to the `[color]`. The GUI can be used to determine which colors are available, and then this command can be used to then chose a particular color (by name). |
| `PyMOL>util.cbag [object]` | This colors the atoms of the `[object]` with carbon = green, oxygen = red, and nitrogen = blue. |
| `PyMOL>show [representation], [selection]` | This shows the representation of the selection. Note that it just adds the representation to shown representations, it does not remove representations. Use the GUI to find the different available representations then use this as a quick method to get back to that representation. |
| `PyMOL>hide [representation], [selection]` | This hides the representation of the molecule. Note that it just removes the one representation. A common command that one might use is `hide everything, [selection]`. This removes all the representations from the active display, giving you a clean slate to work with. |
| `PyMOL>bg_color [representation]` | This sets the background color, and is mostly used to set the background to white for making images for presentations and papers. Many people find a black background more visually appealing and simpler to work with for 'zoomed-in', detailed analysis of a molecular scene (better contrast); a white background is often used at a more global level (at the level of protein chains in an oligomer) and is almost always used for final rendering for purposes of a manuscript, poster, presentation, etc. (less ink used in printing a poster with molecular graphics on white backgrounds). |
| `PyMOL>set [name], [value], [selection]` | This is a subtle and highly flexible command that is can be used to vary literally any of PyMOL's hundreds (to thousands) of parameter settings. Some useful particularly useful settings to consider modifying/customizing are noted below. |
| `PyMOL>set transparency, [value], [selection]` | This adjusts the transparency of any surfaces that are rendered (whether they are actively showing or hidden). The value ranges between 0 (full opacity; the default) and 1 (full transparency). |
| `PyMOL>set surface_color, [color], [selection]` | This changes the color of the surface for the named atom selection. |
| `PyMOL>set sphere_transparency, [value], [selection]` | Same as transparency (above), but adjusts the opacity of any sphere representations, instead of surfaces. |
| `PyMOL>ray [width]` | This initiates *ray-tracing* of the molecular scene that is actively visible in the viewer window, yielding high-quality, photorealistic images. `[width]` specifies the width (in pixels) of the final ray-traced output image (which is written to disk via the 'png' command). |

For more information of molecular visualization and graphics, you can see "An Introduction to Biomolecular Graphics" by Mura *et al.* [1] Note that if any of the results obtained via the procedure described here are included in later work, then the convention is that you will need to cite the software used — e.g., AutoDock Vina, the AutoDock PyMOL plugin, AutoDockTools-4 (which operates behind the scenes in much of what was described above), and PyMOL. The appropriate references are [2, 3, 4, 5].

Your name: ______________________________

For items 1–3, rate each group member (including yourself) on the group evaluation criteria listed below. Use the following scale:

1 = poor | 2 = marginal | 3 = satisfactory/average | 4 = good | 5 = excellent

| Group evaluation criteria | self-evaluation | member 1 name: | member 2 name: | member 3 name: | member 4 name: | member 5 name: |
|---|---|---|---|---|---|---|
| 1. Participated in group meetings | | | | | | |
| 2. Cooperated with group; supported group process | | | | | | |
| 3. Demonstrated consistent commitment and effort | | | | | | |
| | | | | | | |
| List a skill that each student brings to the group (e.g., bioinformatics, writing, bench-work, interpersonal skills, etc.). | | | | | | |
| List a skill which is lacking from your group overall, or which could be improved. | | | | | | |
| List something specific that the group learned from you, that they may not have learned otherwise. | | | | | | |
| Overall, how effectively did your group work together on this task/assignment? | | | | | | |
| Suggest one change the group could make to improve its performance. | | | | | | |

# Known Structure, Unknown Function:
## An Inquiry-based Undergraduate Biochemistry Lab Course

Cynthia Gray, Carol W. Price, Christopher T. Lee, Alison H. Dewald, Matthew A. Cline,
Charles E. McAnany, Linda Columbus, Cameron Mura

## **Supplementary Information, 5**:
## Sample grading rubric from the first term

### Labs 7-9: Recombinant Protein Expression, Chromatography and SDS-PAGE, and Dialysis

**Abstract**

| | | |
|---|---|---|
| Identify Problem Studied | 2.5 pt: | __________ |
| * Isolation of protein | | |
| Mention Techniques Used | 2.5 pt: | __________ |
| * Chromatography, SDS gels | | |
| Relevant Data w/ significance | 2.5 pt: | __________ |
| * Information learned about protein | | |
| Conciseness | 2.5 pt: | __________ |
| (Total: 10 points) | | |

**Introduction**

| | | |
|---|---|---|
| Student understands aims and concepts of the experiment | | |
| Overall Clarity | 4 pt: | __________ |
| Cloning/Expression of Recombinant Protein (8 pts) | | |
| Vector and Antibiotic Selection | 2 pt: | __________ |
| Transformation | 2 pt: | __________ |
| *E. coli* as an expression host | 2 pt: | __________ |
| Induction with arabinose | 2 pt: | __________ |
| Chromatography (16 pts) | | |
| General explanation of chromatography | 4 pt: | __________ |
| Gel filtration | | |
| *Separation based on size | 2 pt: | ___________ |
| *Explanation of method | 2 pt: | ___________ |
| Ion exchange | | |
| *Separation based on pI | 2 pt: | ___________ |
| *Explanation of method | 2 pt: | ___________ |
| Affinity | | |
| *Separation based on a specific interaction | 2 pt: | ___________ |
| *Explanation of method | 2 pt: | ___________ |
| SDS-PAGE (6 pts) | | |
| What does SDS-PAGE do? | 3 pt: | __________ |
| How does SDS-PAGE work? | 3 pt: | __________ |
| Protein of Interest (6 pts) | | |
| Presence and specifics of any affinity tags | 2 pt: | __________ |
| Theoretical MW | 1 pt: | ___________ |
| Theoretical pI | 1 pt: | ___________ |
| Identity | 2 pt: | ___________ |

(Total: 40 points)

**M&M: Student understands experimental design**

Cloning/Expression of Recombinant Protein (3 pts)

| | | |
|---|---|---|
| Plasmid and cell line | 1 pt: | ___________ |
| Media and antibiotic | 1 pt: | ___________ |
| Inducer (IPTG, arabinose, etc.) | 1 pt: | ___________ |

Chromatography (3 pts)

| | | |
|---|---|---|
| Gel Filtration | 1 pt: | ___________ |
| *Sephadex G-100 resin, lysis buffer for elution | | |
| Ion Exchange | 1 pt: | ___________ |
| *DEAE (anion exchange) resin, step elution with increasing [NaCl] | | |
| Affinity | 1 pt: | ___________ |
| *Ni-NTA resin, elute with imidazole | | |

SDS-PAGE (2 pts)

| | | |
|---|---|---|
| *BioRad Ready SDS-PAGE (10-20% Tris-HCl) gel | 2 pt: | ___________ |

Dialysis (2 pt) (Should be in the chromatography section, but optionally can be its own section.)

| | | |
|---|---|---|
| One sentence stating that protein was dialyzed into a new buffer | 2pt: | ___________ |

(Total: 10 points)

**Results: Student understands data processing**

| | | |
|---|---|---|
| Overall Clarity | 2 pt: | ___________ |

Cloning/Expression of Recombinant POI (7 pts)

| | | |
|---|---|---|
| SDS-PAGE gel image | | |
| *Lanes labeled | 1 pt: | ____________ |
| *MW marker labeled | 1 pt: | ____________ |
| *Arrow/circle to indicate induction band (**on each gel!**) | 1 pt: | ____________ |
| Figure caption/legend | 2 pt: | ___________ |
| Text describing results and reference to figure | 2 pt: | ___________ |

Chromatography (21 pts)

*Gel Filtration - 7 pts*

| | | |
|---|---|---|
| SDS-PAGE gel image | | |
| *Lanes labeled (incl. where blue dextran and cyt c were observed) | 1 pt: | ____________ |
| *MW marker labeled | 1 pt: | ____________ |
| Figure caption/legend | 2 pt: | ___________ |
| Text describing results and reference to figure | 3 pt: | ___________ |

*Ion Exchange - 7 pts*

| | | |
|---|---|---|
| SDS-PAGE gel image | | |
| *Lanes labeled | 1 pt: | ____________ |
| *MW marker labeled | 1 pt: | ____________ |
| Figure caption/legend | 2 pt: | ___________ |
| Text describing results and reference to figure | 3 pt: | ___________ |

*Affinity - 7 pts*

SDS-PAGE gel image

| | | |
|---|---|---|
| *Lanes labeled | 1 pt: | ___________ |
| *MW marker labeled | 1 pt: | ___________ |
| Figure caption/legend | 2 pt: | __________ |
| Text describing results and reference to figure | 3 pt: | __________ |
| (Total: 30 points) | | |

**Discussion: Student capable of interpreting data and placing in a broader context**

| | | |
|---|---|---|
| Cloning/Expression results | 3 pt: | __________ |
| *Was transformation and induction successful? How do they know? | | |
| Gel Filtration | 8 pt: | __________ |
| *Did protein elute in expected fraction? Why or why not? | | |
| *Oligomer in void volume | | |
| Ion Exchange | 8 pt: | __________ |
| *Did protein elute in expected fraction? Why or why not? | | |
| *Dicussion of protein pI | | |
| Affinity | 8 pt: | __________ |
| *Did protein elute in expected fraction? Why or why not? | | |
| *Discussion of the His-Tag | | |
| Comparison of the three chromatography methods (yield and purity) | 10 pt: | ___________ |
| Additional citations and outside research | 5 pt: | ___________ |
| *How to improve purification, what can be done after purification | | |
| Error Analysis (That is, they attempt to explain why they did not get the results that they may have expected based on what they know about their protein (oligomeric state, pI, etc.) They don't just say "It didn't work as expected." | 3 pt: | __________ |
| (Total: 45 points) | | |
| **Conclusion**: Summarize Results | 4 pt: | __________ |
| Place results in a broader context | 4 pt: | __________ |
| No introduction of new data/information | 2 pt: | __________ |
| (Total: 10 points) | | |
| **References** | 5 pt: | __________ |

Grand Total of 150 pt: __________

# Known Structure, Unknown Function:
## An Inquiry-based Undergraduate Biochemistry Lab Course

Cynthia Gray, Carol W. Price, Christopher T. Lee, Alison H. Dewald, Matthew A. Cline,
Charles E. McAnany, Linda Columbus, Cameron Mura

## **Supplementary Information, 6**:
## Sample student assessment of their learning gains (SALG) survey questions

### SALG Survey Questions

Instructions to students:

- Teachers value student feedback, which is taken into account when improving courses such as this one. Please be as precise as you can in your answers. Please choose "not applicable" for any activity you did not do. You may find one or more questions at the end of each section that invite an answer in your own words. Please comment candidly, bearing in mind that future students will benefit from your thoughtfulness. Remember that this is an anonymous survey: your teacher will never know what any individual student has written.
- You may see the following note next to some questions:
  "D" — Department question. The department head can view the responses to these questions.

---

## Understanding

1. Presently, I understand...

| 1.1 The following concepts that will be explored in this class | **not applicable** | **not at all** | **just a little** | **somewhat** | **a lot** | **a great deal** |
|---|---|---|---|---|---|---|
| 1.1.1 Literature searches and electronic resources | ○ | ○ | ○ | ○ | ○ | ○ |
| 1.1.2 Reading primary literature | ○ | ○ | ○ | ○ | ○ | ○ |
| 1.1.3 Critiquing primary literature | ○ | ○ | ○ | ○ | ○ | ○ |
| 1.1.4 Writing primary literature | ○ | ○ | ○ | ○ | ○ | ○ |
| 1.1.5 Bioinformatics tools and methods | ○ | ○ | ○ | ○ | ○ | ○ |
| 1.1.6 Molecular visualization | ○ | ○ | ○ | ○ | ○ | ○ |
| 1.1.7 Molecular modeling | ○ | ○ | ○ | ○ | ○ | ○ |

1.1.8 Molecular docking

1.1.9 Buffer solutions

1.1.10 Kinetic assays

1.1.11 Enzyme kinetics

1.1.12 Data analysis

1.1.13 Recombinant protein expression

1.1.14 Chromatography

1.1.15 Protein purification

1.1.16 SDS-PAGE

1.1.17 Dialysis

1.1.18 Protein concentration determination

1.1.19 Protein-ligand binding

1.1.20 Experimental design

1.1.21 Choosing appropriate controls

1.1.22 Systematic perturbation of an experiment to test my hypothesis generated by initial data

1.1.23 Poster preparation

1.1.24 Poster presentation

1.2 The relationships between the concepts listed above

1.3 How ideas we will explore in this class relate to ideas I have encountered in other classes within this subject area

| | | | | | | |
|---|---|---|---|---|---|---|
| 1.4 How ideas we will explore in this class relate to ideas I have encountered in classes outside of this subject area | ○ | ○ | ○ | ○ | ○ | ○ |
| 1.5 How studying this subject helps people address real world issues | ○ | ○ | ○ | ○ | ○ | ○ |

1.6 What do you expect to understand at the end of the class that you do not know now?

---

## Skills

| 2. Presently, I can... | not applicable | not at all | just a little | somewhat | a lot | a great deal |
|---|---|---|---|---|---|---|
| 2.1 Find articles relevant to a particular problem in professional journals or elsewhere | ○ | ○ | ○ | ○ | ○ | ○ |
| 2.2 Identify patterns in data | ○ | ○ | ○ | ○ | ○ | ○ |
| 2.3 Recognize a sound argument and appropriate use of evidence | ○ | ○ | ○ | ○ | ○ | ○ |
| 2.4 Write documents in discipline-appropriate style and format | ○ | ○ | ○ | ○ | ○ | ○ |
| 2.5 Work effectively with others | ○ | ○ | ○ | ○ | ○ | ○ |
| 2.6 Prepare and give oral presentations | ○ | ○ | ○ | ○ | ○ | ○ |

2.7 What do you expect to be able to do at the end of the course that you cannot do now?

2.8 Please comment on how you expect this material to integrate with your career and/or life.

## Attitudes

| 3. Presently, I am... | not applicable | not at all | just a little | somewhat | a lot | a great deal |
|---|---|---|---|---|---|---|
| 3.1 Enthusiastic about the subject | ○ | ○ | ○ | ○ | ○ | ○ |
| 3.2 Interested in taking or planning to take additional classes in this subject | ○ | ○ | ○ | ○ | ○ | ○ |
| 3.3 Confident that I understand the subject | ○ | ○ | ○ | ○ | ○ | ○ |
| 3.4 Willing to seek help from others (teacher, peers, TA) when working on academic problems | ○ | ○ | ○ | ○ | ○ | ○ |

3.5 Please comment on your present level of interest in this subject.

3.6 Why did you choose to take this class?

## Integration of Learning

| 4. Presently, I am in the habit of... | not applicable | not at all | just a little | somewhat | a lot | a great deal |
|---|---|---|---|---|---|---|
| 4.1 Connecting key ideas I learn in class with real research scenarios. | ○ | ○ | ○ | ○ | ○ | ○ |

# Known Structure, Unknown Function:
## An Inquiry-based Undergraduate Biochemistry Lab Course

Cynthia Gray, Carol W. Price, Christopher T. Lee, Alison H. Dewald, Matthew A. Cline, Charles E. McAnany, Linda Columbus, Cameron Mura

## **Supplementary Information, 7**:
## Sample post–course survey questions

Hello: You are invited to participate in this post-course survey to help assess and evaluate Chem4411/21, Biochemistry Labs I & II. The questionnaire should take approximately 20 minutes to complete. Your participation in this study is completely voluntary. There are no foreseeable risks associated with this project. However, if you feel uncomfortable answering any particular questions (unless marked as required), you may skip the question. You can withdraw from the survey at any point. It is very important for us to learn your opinions. Your survey responses will be strictly confidential, and data from this research will be reported only in aggregate. Your information will be coded and will remain confidential. If you have questions at any time about the survey or the procedures, you may contact Cindy Gray by email (cg4eq@virginia.edu). Participation in this survey will enter you into the lottery system described in the preliminary announcement email (we will contact you if you have won a prize from our lottery!). Thank you very much for your time and support. Please start with the survey now by clicking the '**Continue**' button below.

Current Occupation

| |
|---|
| |

For Chem4421 (second semester), who was your Teaching Assistant (TA)?

1. Abelin, Sarah
2. Dawidowski, Alison
3. Ebmeier, Jennifer
4. Fox, Donald
5. Kabzinski, Joseph
6. Kroncke, Ryan
7. Lo, Brett
8. Malaker, Tracy
9. Oliver, Ronald
10. Patterson, Peter
11. Randolph, Jennifer
12. I don't remember

Prior to these courses, did you have any experience in a research-based laboratory?

1. no
2. yes, but only as a course
3. yes, but only as an occupation/internship
4. yes, both as a course and an occupation/internship

What was your overall grade in the courses?

1. A
2. B+
3. B
4. B-
5. C+
6. C
7. C-
8. D+
9. D
10. D-
11. F
12. Prefer Not to Answer

What are your plans, if any, for science education beyond your undergraduate degree?

1. Ph.D. in biology–related field
2. Ph.D. in chemistry–related field
3. Ph.D. in physical science*
4. M.A. in life science*
5. M.A. in physical science*
6. Advanced degree in field other than sciences
7. Medical School (MD)
8. MD/PhD*
9. Other health profession
10. Law or business degree
11. Teaching
12. Peace Corps or similar
13. Work first
14. No school after college, science career
15. No school after college, non-science related career
16. Other

How did the research experience in these courses influence your postgraduate plans?

1. I had a plan for postgraduate education that has not changed.
2. It helped confirm of my postgraduate education consideration.
3. It changed my prior plan in the direction toward a postgraduate education.
4. It changed my prior plan in the direction away from a postgraduate education.
5. I still do not have plans for postgraduate education.

The following statements refer to the poster presentation portion of the courses. Please rate how much you agree/disagree with the following statements:

| | Strongly Disagree | Disagree | Agree | Strongly Agree | Not Sure |
|---|---|---|---|---|---|
| Composing the poster helped me prioritize the data of my research. | ❑ | ❑ | ❑ | ❑ | ❑ |
| Presenting the poster developed my oral scientific communication. | ❑ | ❑ | ❑ | ❑ | ❑ |
| The poster presentations made me more confident in my research. | ❑ | ❑ | ❑ | ❑ | ❑ |
| Overall, the introduction of poster presentation to the courses gave me a deeper understanding of biochemistry. | ❑ | ❑ | ❑ | ❑ | ❑ |

What specific elements of poster presentation did you find useful? What should be improved?

The next few statements refer to the computational aspects of the courses (bioinformatics, databases, literature searches, docking, etc.). Please rate how much you agree/disagree with the following statements:

| | Strongly Disagree | Disagree | Agree | Strongly Agree | Not Sure |
|---|---|---|---|---|---|
| The computational aspects of the courses helped me become more independent in my research. | ❑ | ❑ | ❑ | ❑ | ❑ |
| The computational aspects of the courses made me more confident in my research. | ❑ | ❑ | ❑ | ❑ | ❑ |
| The computational aspects of the courses provided tools for me to be an active participant in discovery. | ❑ | ❑ | ❑ | ❑ | ❑ |
| The computational aspects of the course made my research more tangible. | ❑ | ❑ | ❑ | ❑ | ❑ |
| Overall, I have a deeper understanding of biochemistry due to the computational aspects of these courses. | ❑ | ❑ | ❑ | ❑ | ❑ |

What specific computational aspects of these courses did you find useful? What could be improved?

The next few statements refer to the course lectures/class-times. Please rate how much you agree/disagree with the following statements:

| | Strongly Disagree | Disagree | Agree | Strongly Agree | Not Sure |
|---|---|---|---|---|---|
| I attended lectures regularly. | ❑ | ❑ | ❑ | ❑ | ❑ |
| The lectures worked well in conjunction with the lab. | ❑ | ❑ | ❑ | ❑ | ❑ |
| The lectures were clear and coherent. | ❑ | ❑ | ❑ | ❑ | ❑ |
| The lectures helped me to start thinking independently. | ❑ | ❑ | ❑ | ❑ | ❑ |
| Overall, I have a deeper understanding of biochemistry due to the course lectures. | ❑ | ❑ | ❑ | ❑ | ❑ |

What specific elements of the lectures in the courses did you find useful? What could be improved?

The next few statements will refer to the course labs. Please rate how much you agree/disagree with the following statements:

| | Strongly Disagree | Disagree | Agree | Strongly Agree | Not Sure |
|---|---|---|---|---|---|
| The labs increased my factual knowledge. | ❑ | ❑ | ❑ | ❑ | ❑ |
| The labs increased my critical thinking. | ❑ | ❑ | ❑ | ❑ | ❑ |
| I have retained skills in experimental design because of the labs. | ❑ | ❑ | ❑ | ❑ | ❑ |
| These labs have given me confidence in my research. | ❑ | ❑ | ❑ | ❑ | ❑ |
| I have retained skills in data analysis because of the labs. | ❑ | ❑ | ❑ | ❑ | ❑ |
| I have retained skills in group work because of the lab. | ❑ | ❑ | ❑ | ❑ | ❑ |
| Overall, I have a deeper understanding of biochemistry due to the course labs. | ❑ | ❑ | ❑ | ❑ | ❑ |

What specific elements of the course laboratories did you find useful? What could be improved?

The next few statements refer to manuscript writing. Please rate how much you agree/disagree with the following statements:

| | Strongly Disagree | Disagree | Agree | Strongly Agree | Not Sure |
|---|---|---|---|---|---|
| The courses improved my scientific writing skills. | ❑ | ❑ | ❑ | ❑ | ❑ |
| I learned how to organize my research in a scientific manuscript. | ❑ | ❑ | ❑ | ❑ | ❑ |
| Writing the manuscript gave me more confidence in my research. | ❑ | ❑ | ❑ | ❑ | ❑ |
| Overall, I have a deeper understanding of biochemistry due to the introduction of manuscript writing to these courses. | ❑ | ❑ | ❑ | ❑ | ❑ |

What specific elements of the manuscript writing in the courses did you find useful? What could be improved?

The next few statements refer to group meetings. Please rate how much you agree/disagree with the following statements:

| | Strongly Disagree | Disagree | Agree | Strongly Agree | Not Applicable |
|---|---|---|---|---|---|
| I believe there were a sufficient amount of group meetings. | ❑ | ❑ | ❑ | ❑ | ❑ |
| The group meetings gave me constructive feedback to improve my research. | ❑ | ❑ | ❑ | ❑ | ❑ |
| The group meetings gave me constructive feedback for the final projects of the course (poster and manuscript). | ❑ | ❑ | ❑ | ❑ | ❑ |
| Overall, I have a deeper understanding of biochemistry due to my participation in group meetings. | ❑ | ❑ | ❑ | ❑ | ❑ |

What specific elements of the group meetings in the courses did you find useful? What could be improved?

Please rate how much you agree/disagree with the following statements:

*"Compared to other undergraduate laboratory classes I have taken, this class......*

| | Strongly Disagree | Disagree | Agree | Strongly Agree | Not Sure |
|---|---|---|---|---|---|
| …encourages more independent thinking. | ☐ | ☐ | ☐ | ☐ | ☐ |
| …teaches more skills in time management. | ☐ | ☐ | ☐ | ☐ | ☐ |
| …teaches more skills in scientific communication. | ☐ | ☐ | ☐ | ☐ | ☐ |
| …better prepares students to present scientific information. | ☐ | ☐ | ☐ | ☐ | ☐ |
| …encourages greater confidence in a student's scientific knowledge. | ☐ | ☐ | ☐ | ☐ | ☐ |

Please rate how much you agree or disagree with these statements:

| | Strongly Disagree | Disagree | Agree | Strongly Agree | Not Sure |
|---|---|---|---|---|---|
| I learned to communicate well with my group. | ☐ | ☐ | ☐ | ☐ | ☐ |
| I learned to work professionally with my group. | ☐ | ☐ | ☐ | ☐ | ☐ |
| My group was able to delegate tasks well. | ☐ | ☐ | ☐ | ☐ | ☐ |
| My group met a sufficient amount of time outside of class. | ☐ | ☐ | ☐ | ☐ | ☐ |
| I have a deeper understanding of biochemistry due to working in groups. | ☐ | ☐ | ☐ | ☐ | ☐ |

For the purposes of these courses, I believe the ideal group size would be ______ students.

1. two
2. three
3. four
4. five
5. six
6. seven, or more

Please rank the following items in relation to each other from 1 to 9, with a '1' indicating the greatest contribution to your understanding of biochemistry and a '9' indicating the least contribution.

| | |
|---|---|
| Instructor | ❑ |
| TA | ❑ |
| Group members | ❑ |
| Assigned readings | ❑ |
| Course Lecture | ❑ |
| Course Lab | ❑ |
| Group meetings | ❑ |
| Poster presentation | ❑ |
| Writing a manuscript | ❑ |

Please briefly elaborate on what you ranked in the above list as having the greatest contribution ('1') and what factor you ranked as having the least contribution ('9') to your understanding of biochemistry.

Do you feel that these Chem4411/21 courses adequately prepared you to work more independently in a laboratory setting? If so, how? If not, what would have helped you feel more prepared?

Do you feel that these courses helped you think like a scientist rather than a student? If so, how? If not, how can these courses be improved in that respect?

What are the most positive and most negative differences that you saw in these labs, compared to previous lab courses that you have taken? Please elaborate.

These next few statements will refer to your TA in the Chem4421 course (second semester). Please rate how much you agree/disagree with the following statements:

| | Strongly Disagree | Disagree | Agree | Strongly Disagree | Not Sure |
|---|---|---|---|---|---|
| My TA was able to answer my questions in lab. | ❑ | ❑ | ❑ | ❑ | ❑ |
| My TA was available for questions during their specified office hours. | ❑ | ❑ | ❑ | ❑ | ❑ |
| My TA gave constructive feedback. | ❑ | ❑ | ❑ | ❑ | ❑ |
| My TA was approachable. | ❑ | ❑ | ❑ | ❑ | ❑ |
| Overall, I have a deeper understanding of biochemistry due to the contributions of my TA. | ❑ | ❑ | ❑ | ❑ | ❑ |

What do you think your TA did well? What could your TA improve upon?

Do you have any other comments about these courses?